\documentclass[aps,prb, twocolumn,amsmath,amssymb, superscriptaddress,citeautoscript,floatfix]{revtex4-2}
\usepackage{amsbsy}
\usepackage{wasysym}
\usepackage{bm}
\usepackage{color}
\usepackage{textcomp}
\usepackage{tikz-feynman,contour}
\tikzfeynmanset{compat=1.1.0}
\tikzfeynmanset{/tikzfeynman/momentum/arrow shorten = 0.3}
\tikzfeynmanset{/tikzfeynman/warn luatex = false}
\usepackage{gensymb}
\usepackage[breaklinks,colorlinks = true,linkcolor = red,urlcolor=cyan,citecolor=red]{hyperref}
\usepackage{array}
\usepackage{float}
\usepackage{graphicx}
\usepackage{subfigure}
\usepackage{float}
\usepackage{soul}
\usepackage{appendix}

\usepackage{comment}
\usepackage{dcolumn,multirow}

\begin{document}

\title{
Generation of chirality and orbital magnetization by Stone-Wales-type lattice defects in the Kitaev spin liquid
}

\author{Arnab Seth}
\affiliation{School of Physics, Georgia Institute of Technology, Atlanta, GA 30332, USA}
\author{Fay Borhani}
\affiliation{School of Physics, Georgia Institute of Technology, Atlanta, GA 30332, USA}
\author{Itamar Kimchi}
\affiliation{School of Physics, Georgia Institute of Technology, Atlanta, GA 30332, USA}

\date{February 27, 2026}

\begin{abstract}
In this work we extend our study of the effect of certain crystallographic defects on the spin-1/2 Kitaev honeycomb spin liquid, focusing on its gapless phase and contrasting with the gapped phase. We identify a Stone-Wales (SW) local defect consisting of a 90$^\circ$ bond rotation that preserves Kitaev bond labels for edge-sharing octahedra and thereby enables exact solvability. These SW-type defects involve odd-sided plaquettes with $\pm \pi/2$ fluxes, but can be created locally. An isolated defect hosts a time-reversal pair of ground-state flux configurations with large net chirality. Certain excitations are also chiral. The chirality manifests in Majorana local Chern marker and in scalar spin chirality, producing electronic orbital magnetization. T-matrix analysis and numerics at finite defect density $n_d$ show that defect chiralities generate a topological gap of $11 n_d$ protecting a Chern number $C=\pm 1$. Emergent ferromagnetic long range Ising interactions $r^{-\gamma}$ with $2<\gamma < 3$ between defect chiralities lead to a finite temperature $T_c$ phase transition into the chiral spin liquid.  The $T_c$ is proportional to $n_d$ and diverges when $\gamma\rightarrow 2$. We also consider additional solvable impurity potentials and find that $\gamma$ can be reduced to below $2.3$ and correspondingly enhance $T_c$. Our results offer applications to 2D Dirac cone systems with a finite density of fluctuating Ising magnetic impurities and to identifying spin liquids with lattice defects.

\end{abstract}

\maketitle

\tableofcontents

\section{Introduction}

Incorporating crystalline disorder into theoretical models of strongly correlated systems can impact their expected phase diagrams in unusual ways. 
For example, in 2D spin-1/2 systems, certain quantum paramagnets are unstable to weak bond randomness \cite{kimchi_valence_2018}. Even a slight inhomogeneity in the spin couplings generates an instability, leading to topologically-protected spin defects and closing the gap to spin excitations. For the interesting case of 2D quantum spin liquids (QSLs), including the paradigmatic Kitaev honeycomb QSL \cite{kitaev_anyons_2006, jackeli_mott_2009, savary_quantum_2016,
hermanns_physics_2018, takagi_concept_2019, trebst_kitaev_2022,  matsuda_kitaev_2025},
theoretical treatments have suggested typical stability to weak disorder, but much remains unknown. \cite{willans_disorder_2010,willans_site_2011,sreejith_vacancies_2016, nasu_thermodynamic_2020,kao_vacancy_2023,dhochak_magnetic_2010,kao_disorder_2021,kao_vacancyinduced_2021,singhania_disorder_2023,zschocke_physical_2015,knolle_bonddisordered_2019,
cassella_exact_2023, grushin_amorphous_2023,petrova_unpaired_2014,
sanyal_emergent_2021,halasz_doping_2014,halasz_coherent_2016,
lahtinen_perturbed_2014, 
udagawa_visonmajorana_2018,
kimchi_valence_2018,kimchi_scaling_2018,
otten_dynamical_2019, nasu_spin_2021,freitas_gapless_2022,dantas_disorder_2022,
song_lowenergy_2016,yatsuta_vacancies_2023,vojta_kondo_2016}

Here and in a companion paper \cite{seth_chiral_2025} we show that for the Kitaev honeycomb spin liquid phase, the presence of certain local crystalline defects leads to a finite temperature phase transition into a chiral quantum spin liquid. 
The mechanism for this instability is a local time-reversal-breaking chirality generated by each defect, 
together with an emergent long range interaction between defect chiralities, mediated by the nearly gapless fermion background.
Through this mechanism the instability arises  even at low defect densities.

As a concrete realization of \textit{local} crystallographic defects with the odd-sided plaquettes necessary for breaking time reversal (TR) symmetry through the Kitaev QSL's Majorana fermions \cite{kitaev_anyons_2006}, we here focus on  Stone-Wales-type (SW) type defects well known in honeycomb materials.
\cite{stone_theoretical_1986,vozmediano_gauge_2010, kot_band_2020}. 
The SW defect consists of a pair of pentagons and pair of heptagons, forming a bound state of two 5-7 dislocations with opposite Burgers vectors. 
For isolated 5-7 dislocations \cite{petrova_unpaired_2014}, recent work has shown \cite{borhani_realspace_2025} that in the gapless Kitaev model each constituent  disclination contributes a TR breaking local chirality whose magnitude is complicated but whose sign is universally determined by its flux. (This flux is the emergent flux which couples to the Majorana fermions and arises in the low energy theory of the QSL phase.)
By studying various flux configurations on SW defects below, we show that this result extends to the local SW defect case. The locality of the SW defects also allows us to use the T-matrix formalism to analyze both the sign and the magnitude of the SW chirality contribution and its dependence on the flux configuration.

Indeed, importantly and unlike prior work, 
\cite{yao_exact_2007, dornellas_kitaevheisenberg_2024,peri_nonabelian_2020,cassella_exact_2023, grushin_amorphous_2023,petrova_unpaired_2014, borhani_realspace_2025} 
the SW defects we study are local, i.e.\ can be added to the clean honeycomb Hamiltonian by adding a sum of local impurity potentials. Our results are expected to generalize to other local defects with odd-sided plaquettes, defining a class of SW-type defects. Their locality implies that they can in principle be realized in Kitaev honeycomb materials while preserving the Kitaev bond 3-coloring arising from the Jackeli-Khaliullin mechanism of edge-sharing octahedra \cite{jackeli_mott_2009} as we discuss below.

This manuscript extends the analysis of Ref.~\cite{seth_chiral_2025} by considering:  
(1) the SW flux configurations; 
(2) the two local real space measures of chirality given by Majorana fermion topological orbital magnetization (local marker) in addition to scalar spin chirality; 
(3) numerical computations of both measures, together with a T-matrix analysis, for the  excited state flux configurations as well as the ground state;  
(4) extraction of the anisotropies of the emergent long range interaction between defect chiralities; %
(5) modification of the interaction parameters by three types of local perturbations $\delta H_r$ that can arise from realistic defects, and the resulting changes to the defect-induced chiral QSL instability $T_c$; 
(6) modifications of $T_c$ if defect positions are spatially correlated;
and (7) the case of adding defects to the anisotropy-gapped, in addition to the gapless, Kitaev phase. We also provide additional details for various computations.

\section{Summary of results and outline of this manuscript}

We here give an overview of key results and describe the structure of the remainder of this manuscript. 

We begin in Sec.~\ref{sec_sw} by describing the SW defects. Considering the Jackeli-Khaliullin mechanism of edge sharing octahedra,  we  show that  $\pi/2$ rotation of a hexagonal bond approximately preserves Kitaev bond labels thereby giving an exactly solvable model with local inhomogeneities in the Kitaev spin exchanges and local non-Kitaev perturbations. 
Working to zeroth order in these non-Kitaev perturbations, in Sec.~\ref{sec_fluxes} we describe all the possible flux configurations on the SW defect, and their energetics.

In the two sections that follow, we proceed to show that for an isolated SW defect, all these flux configurations produce corresponding distinct  local chirality breaking time-reversal symmetry. 
We establish this chirality generation in a few ways. In Sec.~\ref{sec_chirality_momentum space} we begin with a T-matrix computation of the local defect potential and the resulting topological gap and its associated global chirality contribution. We also verify it with Chern number computations in momentum space. 
In Sec.~\ref{sec_chirality_real space} we discuss the real space distribution of the  chirality contributed by each flux configuration, which provides additional information. This can be captured by two measures: local Chern marker which gives the Majorana fermion topological orbital magnetization, and scalar spin chirality. Both measures are related to electronic orbital magnetization via charge fluctuations across the Mott insulating gap. 
Though the local marker computation is complicated by the nearly gapless background, we find that its results are similar to the results of scalar spin chirality, and together they provide an understanding of the ground state flux sector chirality as well as the different flux excitations.

We then turn to the case of multiple defects. 
In Sec.~\ref{sec_long-range}, we show that the gapless Majorana fermions mediate a long-range ferromagnetic RKKY-type interaction between these ground state SW chiralities. Representing the defect chirality  by an Ising variable $\mu_r^z$, this long range interaction can be described as
\begin{align}
    &H^{\text{SW}}_{\text{int}}=-\frac{1}{2}\sum_{ r, r'}{J}\left({ r}-{ r'}\right)\mu^z_{ r}\mu^z_{ r'},
    \label{eq_sw_int}
\end{align}
which leads to the alignment of the different SW chiralities in the ground state. We numerically obtain that the interaction follows an approximately power law behavior, $J(r)\approx \left(r_0/r\right)^{\gamma}$, with $\gamma \approx 2.7$ for defect densities $10^{-4}\lesssim n_d \lesssim 10^{-2}$ per site for unperturbed defects.

In Sec.~\ref{sec_tc} we discuss how the long-range interaction leads to a spontaneous breaking of time-reversal symmetry at low temperatures via a true finite temperature phase transition with $T_c / J_K \propto n_d$. Here $J_K$ is the Kitaev interaction defined below and $n_d$ is the defect density as a number of defects per site. The proportionality constant is a $\mathcal{O}(1)$ quantity. Its value is about 2 for unperturbed SW defects but can be  further modified by defect perturbations, for example increasing to 10 for $t_m=0.1$ as described below.  
This $T_c \approx 10 n_d J_k$ is large enough that the $T_c$ can be accessed in experiments for reasonable defect densities and spin interaction strengths. 
The low temperature phase, created by disorder, is the non-Abelian chiral spin liquid. 
Its $T=0$ fractionalization here shows an unusually distinct finite temperature signature, in the form of the orbital magnetization produced by crystallographic defects below $T_c$.

In Sec.~\ref{sec_perturbation} we consider how these results are modified by defect perturbations, focusing on three particular perturbations to the Kitaev exchanges at the defect core. 
These local inhomogeneous perturbations preserve solvability in terms of Majorana fermions enabling our exact computations of their effects.  We show that these perturbations modify the location of the phase boundary between gapless and chiral spin liquid, through a non-universal modification of the $\gamma$ power law of the long range interaction. 
A final Discussion section Sec.~\ref{sec:discussion} summarizes the results and suggests additional open questions raised by considerations of defect perturbations that may be investigated in future work. 

This work focuses on the case of adding defects to the gapless Kitaev QSL. At the end of each relevant section we also consider the comparison to the case of the anisotropy-gapped Kitaev QSL. 
Details are given in Appendices.

\section{\label{sec_sw}Local defects with odd-sided plaquettes and realizable solvable Kitaev couplings }

\begin{figure}
    \centering
    \includegraphics[width=1\linewidth]{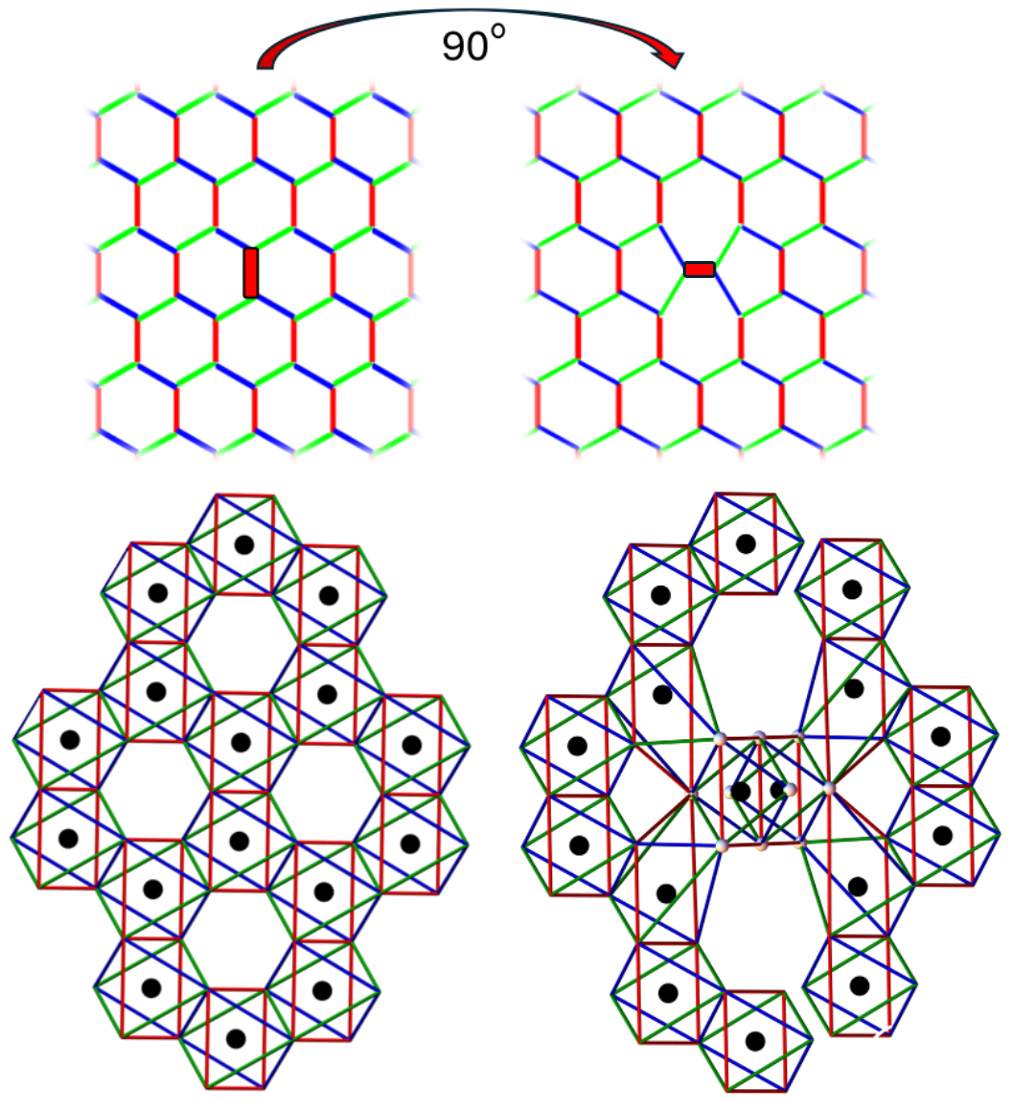}
    \caption{
    {\bf Stone-Wales (SW) type defects in Kitaev honeycomb lattices from edge-sharing octahedra.} Top: a SW defect is created by a $90^\circ$ bond rotation on the honeycomb lattice, with associated modified connectivity. Bond color corresponds to Kitaev interaction axis $x,y,z$. 
    Bottom: in-principle geometric realizability with edge sharing octahedra. Each magnetic site (black disk) is coordinated by six ligands forming the vertices of an octahedron. 
    The color of the edge shared by two octahedra defines the Kitaev interaction axis on the corresponding bond.
    When two octahedra are rotated by  $90^\circ$ around the (0,0,1) axis normal to the red edges, their shared edge remains red and the Kitaev bond axis is exactly preserved.
    The rotation distorts octahedra and produces local out-of-plane buckling, here drawn without any relaxation and depicted via two disconnected octahedra; real material-dependent atomic  relaxation would elastically smooth out the distortions over a few unit cells and generate a material dependent local perturbation $\delta H_r$.}
    \label{fig_SW}
\end{figure}

The Jackeli-Khaliullin mechanism  generating Kitaev exchanges in magnetic Mott insulators stipulates a crystal structure of edge-sharing octahedra \cite{jackeli_mott_2009}.
This mechanism has been extended to other cases, such as metal-organic frameworks where octahedra edges may be connected by other ligands \cite{yamada_designing_2017}, but the geometric considerations are preserved and so for simplicity and without loss of generality we here frame the discussion in terms of edge sharing octahedra. 
 This geometry involves two exchange paths through two different ligands, enabling destructive interference effects that allow anisotropic Ising interactions to dominate, with the Ising axis given by the vector perpendicular to all hopping vectors of the exchange paths. This bond-dependent Ising interaction is exactly the Kitaev interaction.

As shown in Fig.~\ref{fig_SW} the SW defect is created by a 90$^0$ rotation of a honeycomb lattice bond (also see Fig.~1 and main text of the companion paper). 
The Kitaev bond label associated with a shared octahedra edge is invariant under 90$^0$ rotations in the spin-ligand plane, as shown visually by the uniformly-colored squares within each octahedra in Fig.~\ref{fig_SW} bottom left panel. For example, if two octahedra share a red edge, as they do in the bond at the center of the image, then the Kitaev interaction for spins along that bond is colored ``red'' and connects the spin components along the corresponding cubic vector ($z$). Rotating the two central octahedra such that they share another red bond, as in the SW defect image in  Fig.~\ref{fig_SW} bottom right panel, gives the same $z$ direction Kitaev interaction, in addition to other perturbations, thereby also preserving solvability for the Kitaev terms.
These considerations require the 90$^0$ rotation to be in a ``(0,0,1)'' plane that is  tilted relative to the ``(1,1,1)'' honeycomb lattice plane and imply a local out-of-plane distortion for SW defects. Such distortion already arises for Stone-Wales defects in graphene \cite{ma_stonewales_2009}. %
The non-planar distortion enables SW defects to more easily accommodate lattice strain  \cite{peri_nonabelian_2020} or be stabilized by strain \cite{ma_stonewales_2009}. 
Elastic distortions of the octahedra in the surrounding region, which are \textit{not} depicted here,  will smooth out this out-of-plane distortion as well as the other octahedra distortions drawn schematically in Fig.~\ref{fig_SW}.
The SW defects are thus most likely to be stabilized by inter-layer disorder, or by growth over a substrate with impurities; they might also be created by applying strain.

To determine whether a SW or any other defect is indeed stable (or common) for a particular Kitaev-candidate material, and to determine the pattern of elastic distortions which necessarily accompanies the defect, and the resulting effects on the spin Hamiltonian, requires detailed considerations which must be done separately for each specific material of interest. Here we take a broader view of all possible material candidates arising from the Jackeli-Khaliullin mechanism and avoid computing the elastic distortions mentioned above and their  resulting modifications to the electronic Hamiltonian. This also implies that we cannot compute the resulting modifications to the spin exchanges, which we denote as $\delta H_{r}$. However,  even working to zeroth order in $\delta H_{r}$ produces surprising and qualitatively new physical phenomena, including the instability to the chiral spin liquid phase. The resulting gapped spin liquid is expected to be stable to weak perturbations; we also explicitly consider strong perturbations of three particular forms of $\delta H_{r}$ that preserve the spin liquid to compute their effects on the results below. 
We expect that the results will hold for other types of crystallographic defects that modify the honeycomb connectivity, even if the resulting model is no longer exactly solvable.

The modified Kitaev model Hamiltonian in the presence of SW-type defects (i.e., on a modified graph as shown in Fig.~\ref{fig_SW})  can be written as
\begin{align}
    H&=J_K\sum_{\langle ij\rangle}  \sigma^{\alpha_{ij}}_i\sigma^{\alpha_{ij}}_{j}+\sum_{r} \delta H_{r},
    \label{eq_kitaev_1}\\
    \delta H_{r}&=\sum_{\langle ij\rangle \sim r} \delta J_{ij}\sigma_i^{\alpha_{ij}}\sigma_j^{\alpha_{ij}}+\delta H_{r}^\text{NK}
    \label{eq_kitaev_2}
\end{align}
Here the first term is the Kitaev Hamiltonian on the graph with modified connectivity at each SW defect (Fig.~\ref{fig_SW}). 
As usual ${\alpha_{ij}}$ denotes the  three different Kitaev interactions along spin axes $x,y,z$ for bond vector lying on blue, red, green plane respectively, with corresponding bond color in Fig.~\ref{fig_SW}. In this term the interactions are all taken to be of the same magnitude. The term $\delta{H}_{r}$ denote additional  interactions near each SW defect at location $r$, arising due to the quasi-local deformations of the octahedra. 
For the rest of the paper, we work in units such that $J_K=1$ and $a=1$ with $a$ being the nearest neighbor distance on the honeycomb lattice. 

The perturbation term
$\delta H_r$  generically contains two types of interactions as described in Eq.~\ref{eq_kitaev_2}.   
The first term is a Kitaev type bond strength perturbation, which causes bond-strength inhomogeneity in the modified-graph Kitaev exchanges in the vicinity of the defect, which is denoted as $\langle ij\rangle\sim r$. We begin by considering $\delta J_{ij}=0$  which already produces interesting instabilities of the Kitaev model to chiral spin liquid. 
In the last section of this manuscript we will then consider some perturbations within this exactly solvable limit via modifying the Kitaev exchanges on the bonds near the defect, and compute how these perturbations modify the details of the associated phase transition.

The second term in the Eq.~\ref{eq_kitaev_2} represents the non-Kitaev interactions which are perturbations away from the exactly solvable limit. 
Since here we are  concerned with local (non-topological) crystallographic defects, the deformations of the octahedra that generate $\delta{H}_{r}^\text{NK}$ are localized near the defect. 
Therefore, non-Kitaev interactions are generated by the defects only over a  localized region. Because of the dominant Kitaev interaction elsewhere in the lattice and the emergent flux gap, the QSL physics should remain stable to such local perturbations if they are small enough. Their generic effect would then be mild renormalization of the itinerant Majorana density of states and the flux gap.

The SW defects we consider also serve as building blocks for other local defects with 5 or 7 sided plaquettes. Our results for SW defects are expected to generalize to any other local defect that might be more microscopically favorable in any particular Kitaev material.

\section{\label{sec_fluxes} 
Emergent flux configurations and energies}

\subsection{Emergent flux configurations of SW defects}

First let us review and set up notation for the Kitaev model solution in terms of Majorana fermions and gauge field fluxes.
The Kitaev interaction is solved using the Majorana representation of spins: $\sigma_i^\alpha=i b_i^\alpha c_i$, where $c_i$ and $b_i^\alpha$ denote Majorana fermions on the lattice sites. 
This Kitaev spin exchange  Hamiltonian can then be rewritten as
\begin{align}
    H=-\frac{1}{2}\sum_{i,j} i u_{ij}c_ic_j 
    \label{eq_majorana_im}
\end{align}
The factor of $1/2$ appears because each bond is counted twice in the sum. Here $u_{ij}=i b^{\alpha_{ij}}_ib^{\alpha_{ij}}_j$ denotes the emergent $Z_2$ gauge field residing on the bonds of the 3-colored graph. The eigenvalues of this operator are $\pm 1$. This is a conserved quantity as $\left[H , u_{ij}\right]=0$ which gives the exact solvability of the model. 
Note that the above Majorana hopping Hamiltonian can be mapped to  equivalent complex fermion hopping problems which have identical spectra with  doubled degrees of freedom (see Appendix \ref{appen_re-im} for details).  
The  gauge invariant $Z_2$ fluxes can be defined on any closed path (denoted by $\bigcirc$) on the lattice as $W_{\bigcirc}=\prod_{ij\in \bigcirc}\left(-i u_{ij}\right)$. 
It can be shown that the allowed values of these gauge fluxes, in terms of the accumulated phase, are  $\pm 1$ (equivalently denoted as $0$ and $\pi$ fluxes) on any even sided plaquettes; and  $\pm i$ (equivalently denoted as $\pm \pi/2$ fluxes) on any odd sided plaquettes. %

Each SW defect has 4 odd sided plaquettes and thus $2^4=16$ possible flux configurations. Eight of these have  zero net flux. The remaining  eight can be constructed by attaching the net $\pi$ flux on the entire SW defect. 
By incorporating the time reversal ($T$) and bond-centered inversion ($P$) symmetries of a SW  defect, we classify the eight zero-flux configurations into three types, with fluxes $++++, ++--,  +-+-$, in units of $\pi/2$ on the $(5577)$ plaquettes respectively. Each type corresponds to a pair of configurations related by $T$; in addition, the four $+-+-$ configurations come in two pairs related by a lattice mirror symmetry.   
The  $++++$ and $++--$  configurations break $T$ but preserve $P$, while $+-+-$ breaks both $T$ and $P$ but preserves their product $PT$. 

We denote the flux configurations $+-+-$ as PT-flux and $++++$ as Uniform-flux. The $++--$ configuration (and its time reversal partner) will be referred as the Lieb-flux in the rest of the paper. The reason for this nomenclature 
is that their opposite fluxes on 5 and 7 sided plaquettes correspond to a possible extension of Lieb's theorem \cite{lieb_flux_1994}, consistent with prior numerical results in other settings \cite{peri_nonabelian_2020,cassella_exact_2023,petrova_unpaired_2014,borhani_realspace_2025}. 

\subsection{\label{sec_flux energy}Energies: Lieb fluxes as ground states}
We now turn to the flux energetics and find that indeed the Lieb-flux states are the ground states. This motivates additional notation: 
the two fold degeneracy of the Lieb states due to TR symmetry will further be denoted as $\mu^z=+1$ for $++--$ and $\mu^z=-1$ for $--++$ fluxes in units $\pi/2$ on the 5577 plaquettes. This is the notation used for the interaction  Eq.~\ref{eq_sw_int}.

\begin{figure}
    \centering
    \includegraphics[width=0.75\linewidth]{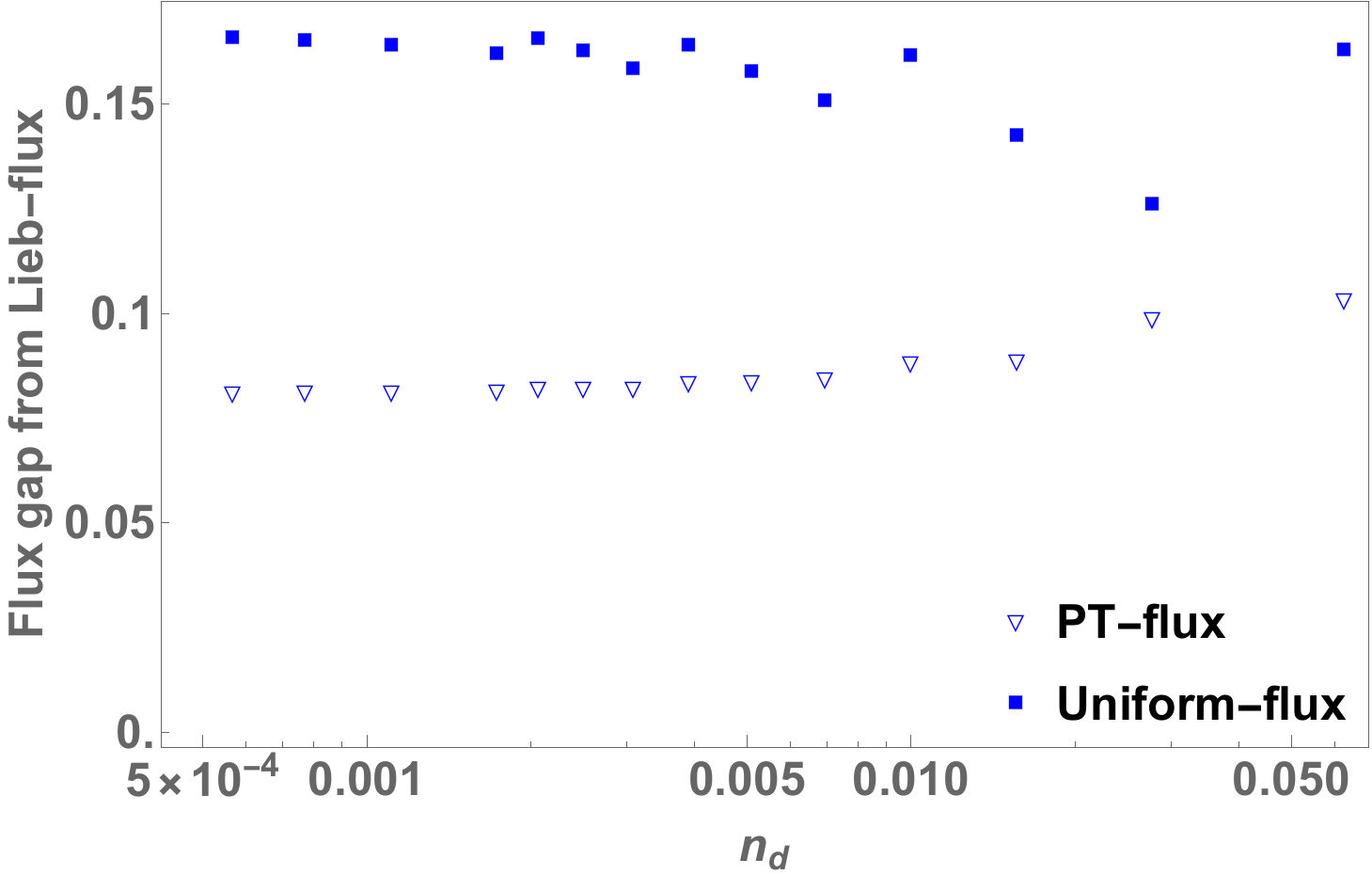}
    \caption{{\bf Flux gap for SW defects at finite density $n_d$.} 
    We compute the total energy of a defect superlattice at density $n_d$ (number of defects per  sites), for the cases where the defects carry Lieb-flux, PT-flux, or Uniform-flux. The Lieb-flux pair of states remain the ground states at all densities, as shown by the flux excitation gaps of the other two states. Flux gap is computed as energy per defect; note the log-linear scale.
    }
    \label{fig_gsenergy}
\end{figure}

We numerically computed the total energy of a system with 3600 sites with an isolated defect and with the defect being in various flux sectors. We also consider the case of two distant defects to consider the possible binding of a net $\pi-$flux on each SW. We find that the Lieb-flux state always remains the lowest energy state among these configurations, while the PT-flux and Uniform-flux states have flux gaps of 0.08 and 0.17  respectively. Recall that we measure energies in units of Kitaev exchange $J_K$.

The $\pi-$flux also remains gapped, though its energy is reduced on a SW defect with a Lieb ground state: binding $\pi-$flux to a 5-sided or 7-sided plaquette on a Lieb-flux SW defect costs energy 0.08 or 0.11 respectively. This energy does not change much with separation of SW defects. Relative to the single-flux $\pi-$flux gap of 0.15 on the clean honeycomb lattice \cite{kitaev_anyons_2006}, this can be interpreted as a $\pi-$flux binding energy of 0.07 on a pentagon within a  Lieb-flux SW defect. Note that since the clean Kitaev model has a local flux gap, associated with the creation of two adjacent $\pi-$fluxes, of 0.26 \cite{panigrahi_analytic_2023}, the corresponding two-$\pi$-flux binding energy of 0.04 is comparable to the SW-defect $\pi-$flux binding of 0.07. 

We further compare the energies by numerically computing them for arrays of  supercells with each supercell hosting a single defect of a particular flux configuration. We find that even in this finite density limit, the Lieb-flux state remains the ground state. This is shown in Fig.~\ref{fig_gsenergy}, where we plot the flux gap of the PT-flux and Uniform-flux states from the Lieb-flux states. In the small density limit, their approximate flux gap is given by  0.08 and 0.16, respectively, which are similar to what we obtained in the isolated defect limit in finite systems. For higher densities the values of the flux gap is modified from the isolated defect limit which can be interpreted as the effect of interaction between the defects.

The two Lieb states  related by the TR symmetry remain the ground states irrespective of the density of SW defects. The ground state flux sector thus involves $\mu^z=1,-1$ on each of the defects. 
However the remaining defect flux states serve as low energy excitations which may be visible in certain experiments. We thus proceed to study all the defect flux sectors in studying the defect generated chiralities, and restrict to the Lieb-flux states only later in this manuscript when we consider the $J(r)$ interactions and low temperature chiral QSL instability. 

Later in this manuscript the TR pair of Lieb states on each defect will be viewed as an Ising spin $\mu^z$; with this language it becomes clear that the other flux states serve as excitations outside of the Ising manifold, and the Ising spins manifold serves as a good low energy theory only for temperatures below the flux excitation gaps.

\subsection{Results for anisotropic gapped Kitaev model}

Note that in the  bond anisotropic gapped Kitaev ``A" phase, the ground state can be obtained by perturbatively computing the energy of the fluxes integrating out the itinerant Majorana fermions. In this phase the  SW defect can be obtained in two different ways, rotating either a strong or a weak bond. 
Following the construction of Section X of  Ref.~\cite{petrova_unpaired_2014}, it can be shown that two Lieb-flux states with $\mu^z=\pm 1$ are the ground states, with their stabilization energies being sensitive to the nature of the rotated bond (see  Appendix \ref{appen_gapped kitaev} for details). In the  scenario of the gapless QSL, this approach does not apply and we have to rely on the numerical computations.

\section{\label{sec_chirality_momentum space}Chirality generation by defect fluxes: T-matrix and Chern number}

\begin{figure}[t]
    \centering
    \includegraphics[width=1.\linewidth]{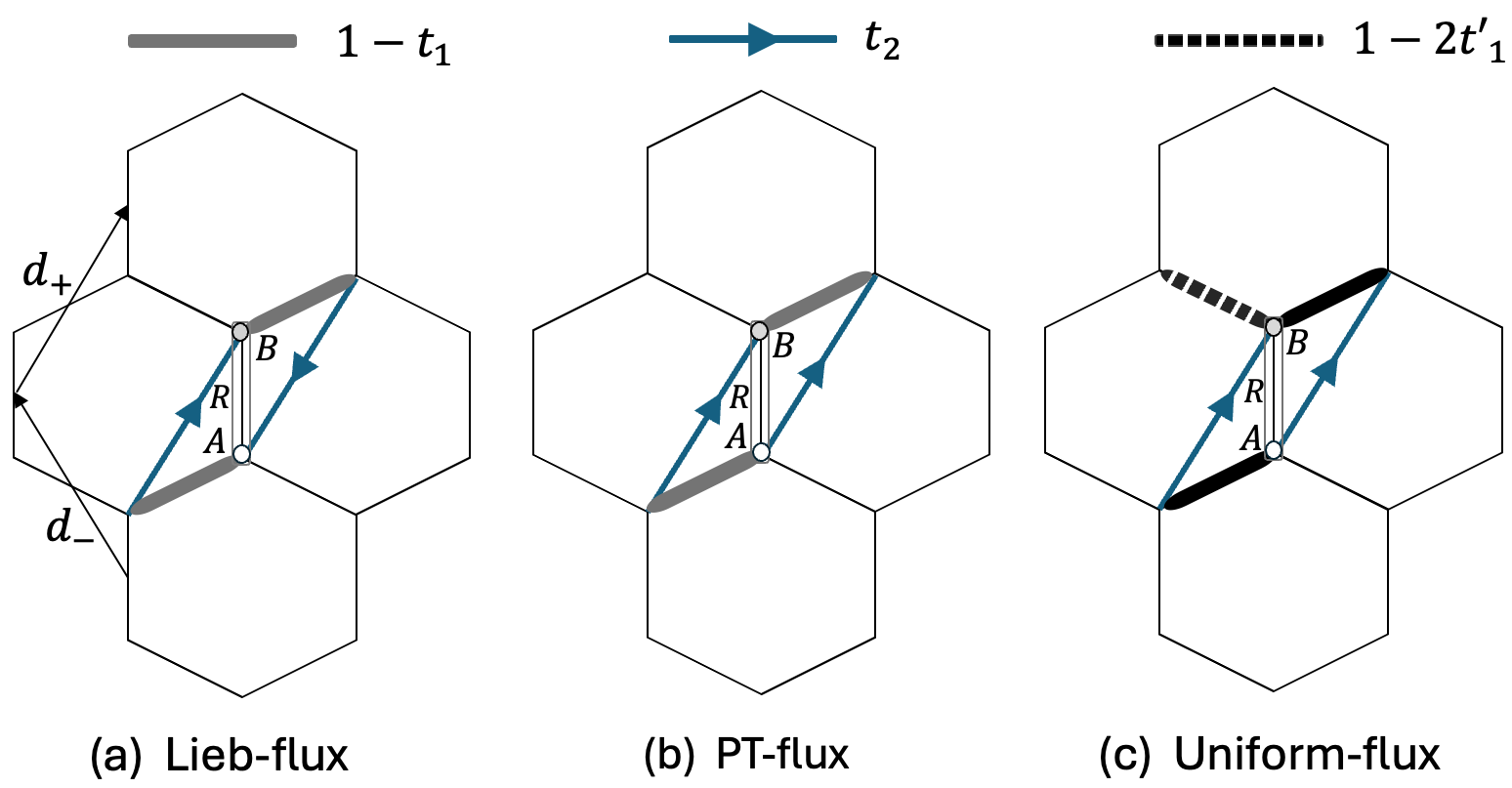}
    \caption{{\bf Creating SW defects by  adding Majorana hopping terms.} %
    Here we consider a representation of a SW defect which consists of adding terms to the quadratic Majorana Hamiltonian $H_0$ of the Kitaev honeycomb model. We consider a SW defect arising from  $+\pi/2$ rotation of the central bond, for each of the three flux configurations (panels a,b,c respectively). The modified bonds are highlighted; other bonds remain the same as in $H_0$ (Eq.~\ref{eq:majorana_im_H0}), with magnitude 1 (multiplying the imaginary $i$).
    For each modified bond, the resulting modified hopping amplitudes of the full Hamiltonian $H'$ are shown at the top of the figure. For the nearest neighbor bonds these amplitudes ($1-t_1$ and, for (c), $1-2t_1'$) give the bond hopping multiplying the imaginary unit as in $H_0$, so that $t_1=1$ removes the corresponding bonds, and $t'_1=1$ flips the corresponding bonds (of uniform-flux SW). For the second neighbor bonds ($t_2$) the hopping is $i t_2$ in the direction shown by the arrow.  The parameters  $t_1$, $t_1'$ and $t_2$ are zero for the honeycomb model, become positive upon perturbative addition of a defect, and take the value $1$ for the fully-added SW defect.  Generic values of $t_j\neq 0,1$ spoil the solvable three-coloring, which is restored  in the $t_j=1$ limit via removal of the $1-t_1$ bonds. 
    The perturbations shown here generate the same fluxes as shown in Figs.~\ref{fig_BiancoResta_SW} and  \ref{fig_uniform}.
    We highlight the unit cell associated with the central bond at $R$ and label its sublattices $A,B$; Bravais lattice vectors $d_\pm$ are also labeled in panel (a).}
    \label{fig_flux_perturbations}
\end{figure}

\subsection{T-matrix analysis: set up}

In this section, we study the chirality generation by an isolated SW defect from a perturbative point of view. This treatment is analogous to an impurity potential problem in a clean free fermion hopping model which is enabled due to the locality of the SW defects.  We also perform resummed perturbation theory using the T-matrix formalism. The results for Lieb flux is also given in the companion  paper~\cite{seth_chiral_2025}. Here we give the results for other flux sectors as well which show additional interesting chirality effects that can give rise to potential experimental signatures at finite temperatures. 

The SW defects can be expressed as a local perturbation $V^{\rm SW}$ added to the gapless Kitaev Hamiltonian on the clean honeycomb lattice $H_0$. This is illustrated in Fig.~\ref{fig_flux_perturbations} in a particular gauge choice. Denoting the clean Kitaev honeycomb model as $H_0$, we can add SW defects by adding the additional terms $V^{\rm SW}$, as
\begin{align}
    H'=H_0+V^{\rm SW}.
    \label{eq_h0+v}
\end{align}

Recall that using the Majorana decomposition of the spin given in Sec \ref{sec_fluxes}A,  $H_0$ in the ground state zero flux sector (and in the usual gauge choice) can be written as 
\begin{align}
    H_0=\sum_{R,\ \nu=0,\pm}  ic_{R,A}c_{R-d_\nu,B},
    \label{eq:majorana_im_H0}
\end{align}
where $c_{R',S}=c_{R',S}^\dagger$ is a Majorana fermion  on sublattice  $S=A,B$ of the unit cell at  $R'$. The $d_0=0$ and $d_\pm=\frac{\sqrt{3}}{2}\left(\pm 1, \sqrt{3}\right)$ are the Bravais lattice vectors of the honeycomb model. The sum over $R$ is over unit cells. The sum over $\nu$ is over $0,+,-$.   
This is the same as 
Eq.~\ref{eq_majorana_im} of the clean honeycomb model with gauge choice of $u_{ij}$ being $+1$ when sublattices  are $i\in A$, $j\in B$.

Now we add the defect.
To obtain definite SW flux sectors of  Lieb, PT and Uniform-flux configurations, we write the perturbations $V^{\rm SW}$ as
\begin{align}
    &V^\text{SW}_\text{Lieb}= \left(c_{R,A} , c_{R,B}\right)\left(
t_1 \sigma^y - i  t_2\mu^z \sigma^0
\right)\left(
    \begin{array}{c}
         c_{R+d_+,A}  \\
         c_{R-d_+,B} 
    \end{array}
    \right),\\
    &V^\text{SW}_\text{PT}= \left(c_{R,A} , c_{R,B}\right)\left(
t_1 \sigma^y + i  t_2\mu^z_\text{PT} \sigma^z
\right)\left(
    \begin{array}{c}
         c_{R+d_+,A}  \\
         c_{R-d_+,B} 
    \end{array}
    \right),\\
    &V^\text{SW}_\text{Uniform}= \left(c_{R,A} , c_{R,B}\right)\left(
t_1 \sigma^y 
+ i  t_2\mu^z_\text{U} \sigma^z
\right)\left(
    \begin{array}{c}
         c_{R+d_+,A}  \\
         c_{R-d_+,B} 
    \end{array}
    \right)\nonumber\\
    &\hspace{4.5cm}+ 2i t_1'  c_{R,B}c_{R+d_-,A},
\end{align}
respectively. Here $\sigma$s are Pauli matrices acting on sublattice indices and $\sigma^0$ is the identity matrix.
The bond midpoint $R$ is the SW inversion center.
We always consider the perturbation parameters $t_1, t_1'$ and $t_2$ to be positive when the SW defect is added. $H'$ represent the actual SW defect full Hamiltonian $H$ of Eq.~\ref{eq_majorana_im} in the limit of $t_1=t_1'=t_2=1$.

All these expressions have two types of terms. The terms with $t_1$ and $t_1'$ are TR symmetric bipartite-preserving perturbations which  perturbatively weaken  two first nearest neighbor bonds connected to the SW central bond. 
For generic values of $t_1$, the perturbation violates the three-bond connectivity of the Kitaev model, which is restored for $t_1=1$ with complete removal of two nearest neighbor bonds. To obtain the physical SW flux configurations, we need to set $t_1=t_1'=1$. In this limit $t_1'$ flips the sign of the hopping on this particular A-B bond, and adds $\pi$ fluxes to the two plaquettes sharing that bond.

The terms with $t_2$ coefficient break TR and are not bipartite in that they connect sites $A$ to $A$ and  sites $B$ to $B$. We will refer to such perturbationas that create AA or BB bonds as ``non-bipartite''. Flipping the overall sign of these terms flips the net TR breaking effect, which is captured here via the Ising variables $\mu^z, \mu^z_\text{PT}, \mu^z_\text{U}=\pm1 $. Convention for $\mu^z$ is already described at the beginning of Sec. \ref{sec_flux energy}. We define $\mu^z_{\rm  PT}=+1$ and $\mu^z_{\rm U}=+1$ for the flux configurations as shown in Fig.~\ref{fig_BiancoResta_SW}(b) and \ref{fig_uniform}(a), respectively (also obtained in Fig.~\ref{fig_flux_perturbations} in the physical SW limit of  $t_1=t_1'=1$).

The choice of impurity potential involves a type of gauge freedom.
In this particular gauge choice described in Fig.~\ref{fig_flux_perturbations}, $V^\text{SW}_\text{Lieb}$ and $V^\text{SW}_{\rm PT}$ preserves the symmetries of the exact Lieb and PT-flux state. However, $V^\text{SW}_\text{Uniform}$ breaks additional inversion symmetry which is not broken by the Uniform-flux configuration itself. This can be correctly restored if the inversion is projectively implemented via suitable gauge transformations.

\subsection{T-matrix analysis: first order computation}

To the first order in perturbation theory, we now compute the effective low energy Hamiltonian due to the above interactions acting on the low energy sector of the pristine Kitaev honeycomb model. Note that in the IR limit, the pristine Kitaev honeycomb model is effectively described by two Majorana Dirac cones:
\begin{align}
    & P[ H_0 ] = v_F\sum_{ q} {\psi}^{\dagger}_{ q}\left(q_y\sigma^x-q_x\tau^z\sigma^y\right){\psi}_{ q}
\label{eq_sw perturbation}
\end{align}
where $P$ symbolically denotes the low energy projector, 
${\psi}^{\dagger}_{ q}=\left(c_{{ K+q},A}~~c_{{ K+q},B}~~c_{{ K'+q},A}~~c_{{ K'+q},B}\right)$ with ${ K}$ and ${ K'}$ being the position of the two Dirac cones, $v_F=3/2$ is the Fermi velocity of the Majorana cones, and $\sigma$s and $\tau$s are Pauli matrices acting on the sublattice and valley, respectively. 
Projecting the defect interactions in this low energy theory, we obtain:
\begin{widetext}
\begin{align}
   &P[V_\text{Lieb}^\text{SW}]=\frac{1}{\mathcal{N}_c}\psi^\dagger_q\left(\sqrt{3}\mu^zt_2 \tau^z\sigma^z-t_1 \vec{m}_\text{t}\cdot(\tau^x, \tau^y)\sigma^y-2t_1\vec{\phi}\cdot (\sigma^y,\sigma^x\tau^z)\right)\psi_q\\
    &P[V_\text{PT}^\text{SW}]=\frac{1}{\mathcal{N}_c}\psi^\dagger_q\left(\sqrt{3}\mu^z_\text{PT}t_2 \tau^z-t_1    \vec{m}_\text{t}\cdot(\tau^x, \tau^y)\sigma^y-2t_1\vec{\phi}\cdot (\sigma^y,\sigma^x\tau^z)\right)\psi_q\\
    &P[V_\text{Uniform}^\text{SW}]=\frac{1}{\mathcal{N}_c}\psi^\dagger_q\left(\sqrt{3}\mu^z_\text{U}t_2 \tau^z-t_1    \vec{m}_t\cdot(\tau^x,\tau^y)\sigma^y-t_1'    \vec{m}'_t\cdot(\tau^x,\tau^y)\sigma^y-(2t_1\vec{\phi}+t_1'\vec{\phi}^*)\cdot (\sigma^y,\sigma^x\tau^z)
    \right)\psi_q
\label{eq_1st order perturbation}
\end{align}    
\end{widetext}
where $\mathcal{N}_c$ denotes the  number of unit cells, $\vec{\phi}$ is a vector made out of real and imaginary components of  $e^{i\pi/3}$, $\vec{\phi}^*$ denotes its complex conjugation. This phase factor is sensitive to whether the rotation of the central bond (here a $z$ bond of Kitaev model) is clockwise or counterclockwise to obtain SW. Here we give the result following the counterclockwise rotation as described in Fig.~\ref{fig_flux_perturbations}. Following the symmetry transformation properties of the low energy operators given in Appendix \ref{appen_symmetry}, we can show that this term preserves time-reversal and inversion, hence merely shifts the Dirac cone in an inversion symmetric manner without opening a gap. 
In the second terms, the coefficients  $\vec{m}_\text{t}(R)$ and $\vec{m}'_\text{t}(R)$ denote  two component vectors consisting of  real and imaginary part of $\text{exp}\left(i(K-K')\cdot R\right)$ and $\text{exp}\left(i(K-K')\cdot R+i\pi/3\right)$, respectively.  These terms are again a TR preserving term but implement the translation symmetry breaking effect via the defects. They anticommute with the free Dirac matrices, hence act on the two Dirac cones as trivial mass terms. 
However, due to the defect position dependence of this term, it vanishes when averaged over a realistic random defect configuration. In fact such a term  vanishes even when a single defect is considered in a finite open system, due to the interference effects from the boundaries \cite{xu_chirality_2025}.

Turning to the term with $t_2$ coefficient, we first note that it breaks TR for all three cases. It preserves inversion for Lieb; although for Uniform-flux the inversion symmetry is preserved, the gauge does not respect it. For the Lieb-states, the corresponding term ($\tau^z\sigma^z$) anticommutes with kinetic Dirac matrices, hence gives rise to topological mass terms. At the first order perturbation theory, this mass dominates over other trivial mass term in the above expression leading to the origin of non-trivial topology. Note that the sign of the chirality is controlled by the TR breaking Lieb configuration set by $\mu^z$. Following the imaginary-bond Haldane model of a Kitaev model in a magnetic field \cite{kitaev_anyons_2006}, which shows similar terms in the low energy, a $\tau^z\sigma^z$ term with positive coefficient  generates a positive (anticlockwise) chirality, which for a gapped bulk gives Chern number $C=1$.
For PT and Uniform-fluxes, the $\tau^z$ term cannot open the gap in the Dirac cones because it does not give sublattice imbalances. Rather it gives tilting of the two cones because of inversion symmetry breaking. 

However, it is generically expected that this first order perturbative results get modified when the full re-summed theory is considered. Also note that the physical SW limit of $t_1=t_1'=1$ may not be captured correctly by the mere first order perturbation. Therefore, we turn to look at the complete T-matrix due to the  interactions which takes care of the scattering from the defect at all orders.

\subsection{\label{sec_tmatrix}
T-matrix analysis: T-matrix results including infinitely many scattering events}

The T-matrix for a defect potential $V$ located at $R$ is given by,
\begin{align}
    T(E)=V (1-G_0(E) V)^{-1}
\end{align}
where $G_0(E)$ represent the Green's function of the unperturbed Hamiltonian at energy $E$. We further project  to the low energy to Dirac cones ($T(E=0)$) to obtain the effective IR contributions. For different fluxes, these are given by
\begin{widetext}
\begin{align}
    & P[T_\text{Lieb}^\text{SW}]=\frac{1}{\mathcal{N}_c}f_1 \psi_q^\dagger\left(a_1 t_2\mu^z\tau^z\sigma^z-b_1 \vec{m}_t\cdot \left(\tau^x,\tau^y\right)\sigma^y-c_1\sigma^y-d_1\sigma^x\tau^z
    \right)\psi_q
    \label{eq_tmatrix_lieb}\\
    &P[T_\text{PT}^\text{SW}]=\frac{1}{\mathcal{N}_c}f_2 \psi_q^\dagger\left(a_2 t_2 \mu^z_\text{PT}\tau^z-b_2 \vec{m}_t\cdot\left(\tau^x, \tau^y\right)\sigma^y-c_2\sigma^y-d_2\sigma^x\tau^z
    \right)\psi_q
    \label{eq_tmatrix_pt}\\
    &P[T_\text{Uniform}^\text{SW}]=\frac{1}{\mathcal{N}_c}f_3 \psi_q^\dagger\left(a_3 t_2 \mu^z_\text{U}\tau^z\sigma^z+a_3't_2 \mu^z_\text{U}\tau^z-\vec{m}_t\cdot\left(b_3\tau^x,b'_3\tau^y\right)\sigma^y-c_3\sigma^y-d_3\sigma^x\tau^z
    \right)\psi_q
    \label{eq_tmatrix_uniform}
\end{align}

\end{widetext}
Detailed expressions of the coefficients are in  Appendix \ref{appen_tmatrix}.
Setting the physical SW limit of $t_1=t_1'=1$ we see  that both the topological and trivial gaps for Lieb-flux  survive even after  resumming the T-matrix. The trivial mass term is generically expected to vanish when open boundary is concerned due to interference from the boundaries or in a realistic uncorrelated disorder configuration due to the disorder averaging. Therefore, the topological term always contributes to the non-zero chirality generation. Further note that  this topological mass term never changes sign with tuning $t_2$, leading to fixed chirality states set by the physical $\mu^z$. 

For the PT states (with $t_1=1$ in $P[T^\text{SW}_\text{PT}]$), T-matrix does not generate any topological mass, so this flux can not generate any net chirality. However, in the following sections, we will see that its individual disinclination can still contribute to individual chirality which does not vanish at the lattice length scales, and produce a chirality pattern with zero net circulation.

For Uniform-flux states (with $t_1=t_1'=1$ in $P[T^\text{SW}_\text{Uniform}]$), the T-matrix produces several interesting effects.
In the physical SW limit, the inversion symmetry breaking  $a'_3 \tau^z$ term vanishes, thanks to the T-matrix resummation. Thus the resummation restores the inversion symmetry of the underlying flux pattern (which was not manifest in the first-order term). 

Remarkably, the Uniform-flux resummed T-matrix also generates a new topological mass term  $a_3 \tau^z \sigma^z$  which was not present in the first order perturbation theory. This indicates that this flux state also contributes to a nonzero chirality.

Even more interestingly, the coefficient of this topological term remains negative  for small $t_2$ similar to the Lieb state, but then flips sign at 
\begin{align}
t_{2,\text{chirality-reversal}}=\sqrt{\frac{2\pi+3\sqrt{3}}{4\pi-3\sqrt{3}}}\approx1.25
\end{align}
indicating a chirality reversal.
Similar defect driven chirality reversing transition in Dirac systems has been observed recently in other settings \cite{neehus_genuine_2025,xu_chirality_2025}.
This has interesting implications. In real candidate Kitaev materials hosting SW defects, $t_2$ is determined by the bond lengths of  lattice as well as by the details of geometrical deformation to accommodate the defect. Therefore, it can vary across materials. Depending on this microscopic parameter, the Uniform-flux state would then give rise to different Chern numbers. 
As we shall see below, the uniform-flux defects can also produce different Chern numbers depending on their density.

\subsection{Numerically computed gap and Chern number for defect superlattice}

\begin{figure}[t]
    \centering
    \includegraphics[width=1.\linewidth]{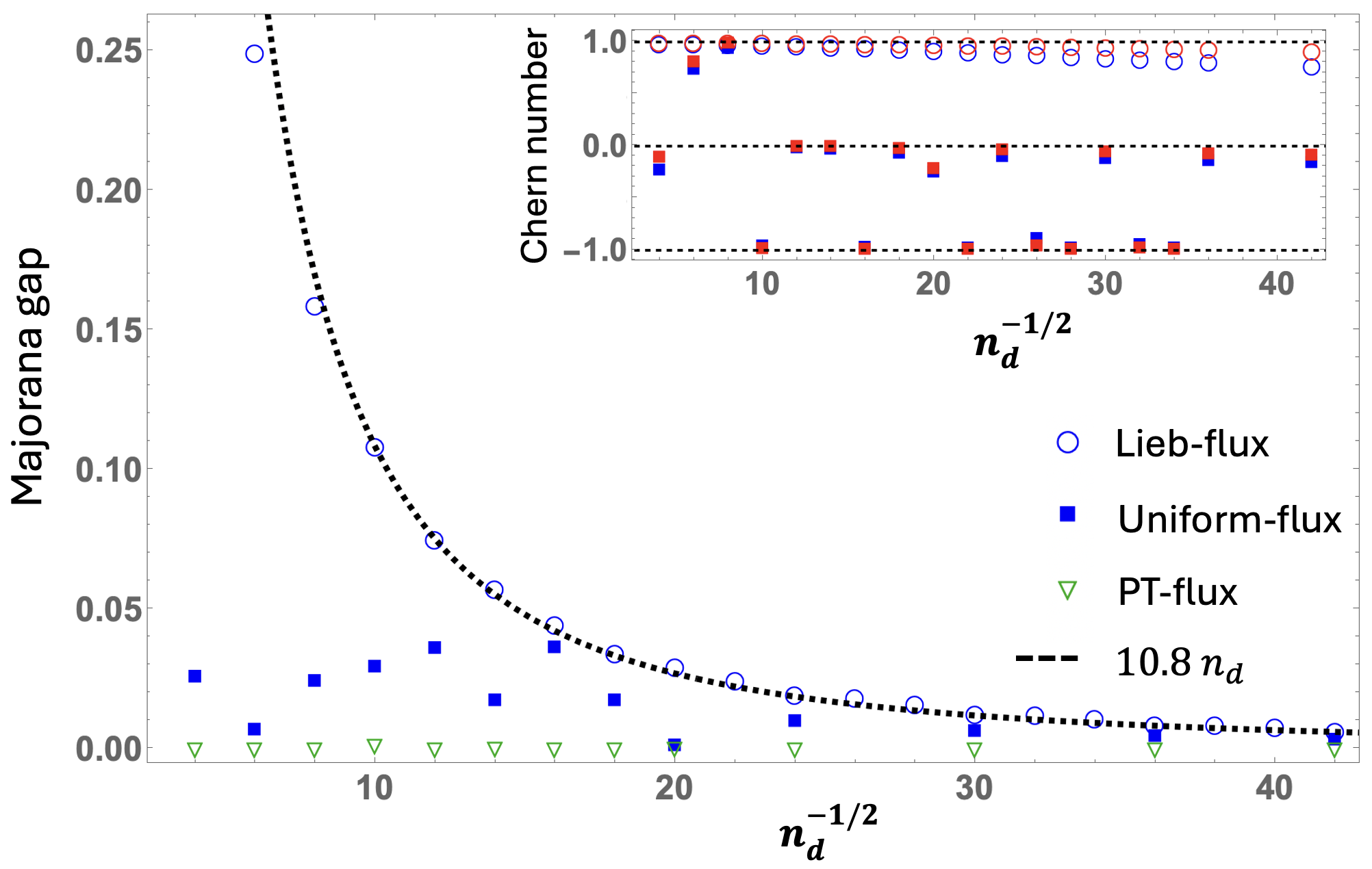}
    \caption{{\bf Majorana gap and Chern number for different fluxes in array of SW defects.} 
    In these computations of the Kitaev model an array of defects with defect density $n_d$ is obtained by placing one SW defect per unit cell of dimensionless linear size $l_u=n_d^{-1/2}$.
    Main panel shows Majorana fermion gap.
    The Lieb states (circles) always show a finite gap, scaling as $10.8 n_d$ (dashed line). The PT-fluxes (triangles) remain gapless for all $n_d$. The gap for the Uniform-flux defects (squares) behaves in a complicated non-monotonic manner and can become very small for some densities.
    The inset shows the  Chern number $C$ for Lieb and Uniform-fluxes (circles and solid squares respectively). 
    Even for small finite gaps the integrated Berry curvature converges to the quantized Chern number with increasing integration mesh ($36\times 36$,  $72\times72$ in blue, red respectively). %
    The sign of the Chern number is determined by the imaginary flux configuration: here Lieb-flux defects are taken with $\mu^z=1$ (5577: $++--$ in units of $\pi/2$) and show $C=1$. Uniform-flux defects are taken with $\mu^z_U=-1$ (fluxes $++++$ in units of $\pi/2$) and show $C$ taking any of the three values $-1,0,1$.
    Note that Majorana gap and Chern number for the Lieb-fluxes are also shown in Ref.~\cite{seth_chiral_2025} Fig.~2. 
    }
    \label{fig_majorana gap_chern}
\end{figure}

To complement our T-matrix computation, here we consider a supercell periodic arrangement of defects and compute the Chern number of the system which represents the chirality for the superlattice  translation invariant systems. We place a single defect in a  supercell with $\ell_u^2$ honeycomb lattice points, hence effective defect density here is $n_d=\ell_u^{-2}$. Depending on microscopic mechanism for defect formation, such as growth on a substrate inducing periodic out of plane buckling, such a spatially correlated defect array may also be physically relevant. We show the Majorana gap and Chern number of such defect configurations for conventional SW defects ($t_2=1$) in Fig.~\ref{fig_majorana gap_chern}.

The Majorana gap for Lieb-flux state remains always non-zero and scales with defect densities as $\approx10.8n_d$ (also shown in Fig.~2 of the companion paper). Computing the Chern number of the system, we find that it is always quantized to $\pm 1$ for $\mu^z=\pm1$. We  further check that the Chern number does not change with tuning  $t_2$, which is consistent with T-matrix results.

The PT-flux states always remains gapless for all defect densities, hence Chern number is not defined in this case. 

For the Uniform-flux case, the Majorana gap behaves in a complicated manner showing either gapless or gapped behavior depending on the densities. For large densities,  the gap is robust and the Chern number quantizes to $\pm1$ for $\mu^z_U=\mp1$. The T-matrix  description developed for an isolated defect is not  adequate to explain this finite density observation, but the sign of the non-zero Chern number in the low densities is consistent with the chirality obtained from the T-matrix at physical SW limit.

\begin{figure*}[tb]
    \centering
    \includegraphics[width=1.8\columnwidth]{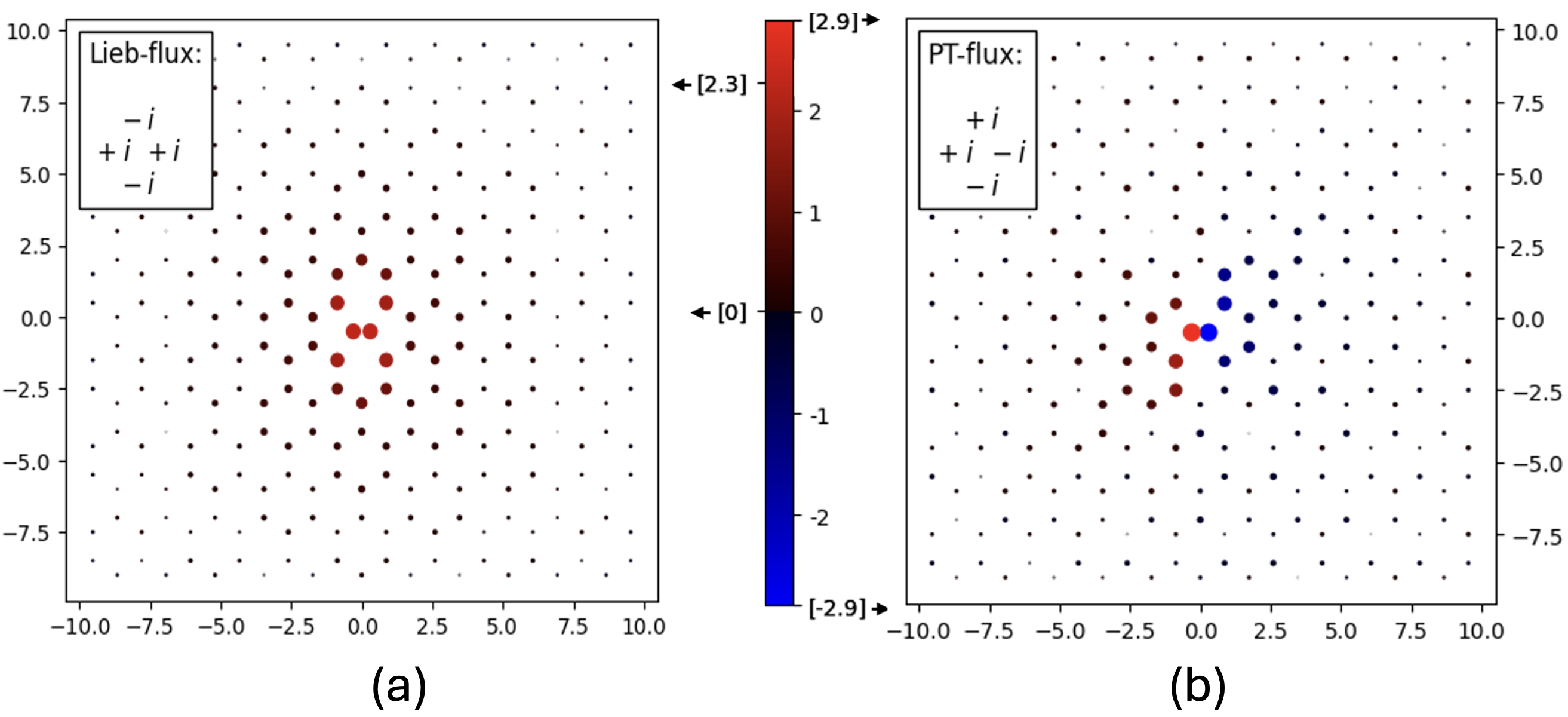}
    \caption{ {\bf{Local marker contribution of defects to chirality: }} Local Chern marker pattern generated for
    (a) Lieb-flux, and (b) PT-flux  with 
    sign of fluxes corresponding to $\mu^z=1$ and $\mu^z_\text{U}=1$, respectively. 
    Local marker on each site is shown as disk with corresponding area. Max/min values are shown in brackets on color scale. The marker is largest near the defect core and decays away from it.  For (a) the contribution is always positive in the bulk, while for (b) it is symmetrically distributed about zero.
     These local marker patterns can be viewed as arising from monopolar (a) and  dipolar (b) distributions of     chirality over the constituent disclinations.
    }
    \label{fig_BiancoResta_SW}
\end{figure*}

\section{\label{sec_chirality_real space}Chirality contributions in real space}

To further describe the chirality induced by the defects, here we discuss and compute two  measures of chirality: scalar spin chirality (SSC) and local marker corresponding to  Majorana fermion topological orbital magnetization. We will use a local description of these observables since we are interested in defect contributions. The local marker has been previously computed for odd sided plaquettes in Kitaev models \cite{cassella_exact_2023,grushin_amorphous_2023,borhani_realspace_2025}, as has scalar spin chirality \cite{grushin_amorphous_2023}, and both are known to peak locally near defects \cite{grushin_amorphous_2023,borhani_realspace_2025}.

The quantities are nominally defined at the lattice scale. 
Physical observables will be coarse grained.
However the local marker in particular requires a fully gapped system to be well defined. The present case of nearly gapless Majorana fermion background produces complications. Note that the nearly gapless background can be viewed as the gapless Dirac cone of the clean system with additional local Dirac mass terms. These local mass terms are known to generate a spectral gap even globally (e.g.\ see Supplementary III of Ref.~\cite{neehus_genuine_2025}) but this is global gap is very small and proportional to defect density, hence the background is nearly gapless. 
This point is also related to the discussion in  Sec.~VII(C) below, and to the result of Sec.~VIII that global chirality (hence global gap) is indeed generated at low temperatures, though the results of the present Section VI will show that it is primarily concentrated near defects. 

We use a local marker $M_1$ that modifies  the conventional local marker $M_0$ for contributions arising near the boundary, thereby more effectively constraining contributions to a bulk region, to minimize this issue. Comparing to the scalar spin chirality, which has no ambiguities in its lattice scale definition, we find that both measures show the same qualitative features such as monopole/dipole/quadrupole chirality patterns. We thus gain confidence that these features will remain visible upon coarse graining when measured in experiment.

\begin{figure*}[t]
    \centering
    \subfigure[]{\includegraphics[width=0.6\columnwidth]{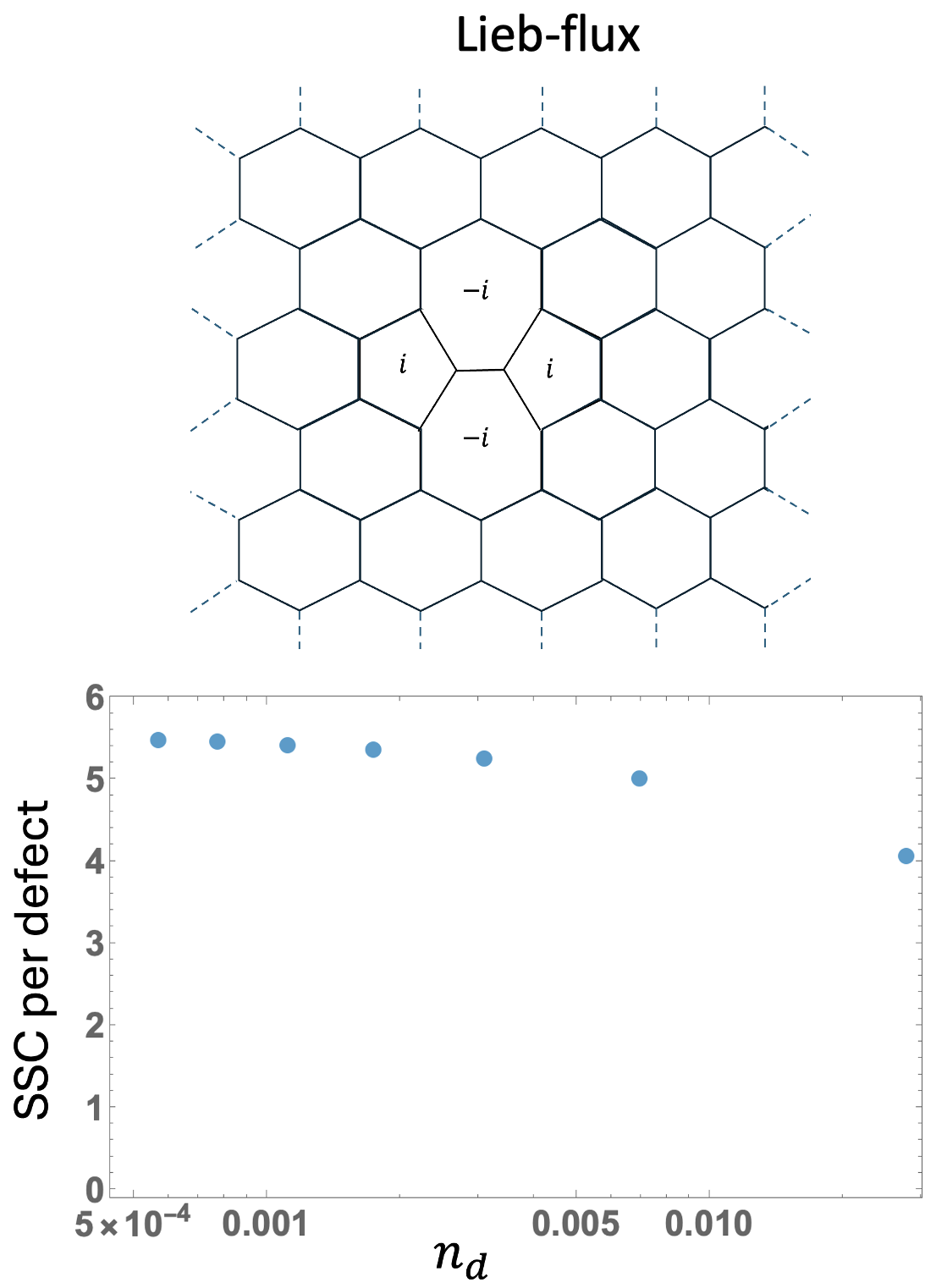}}
    \subfigure[]{
    \includegraphics[width=0.3\linewidth]{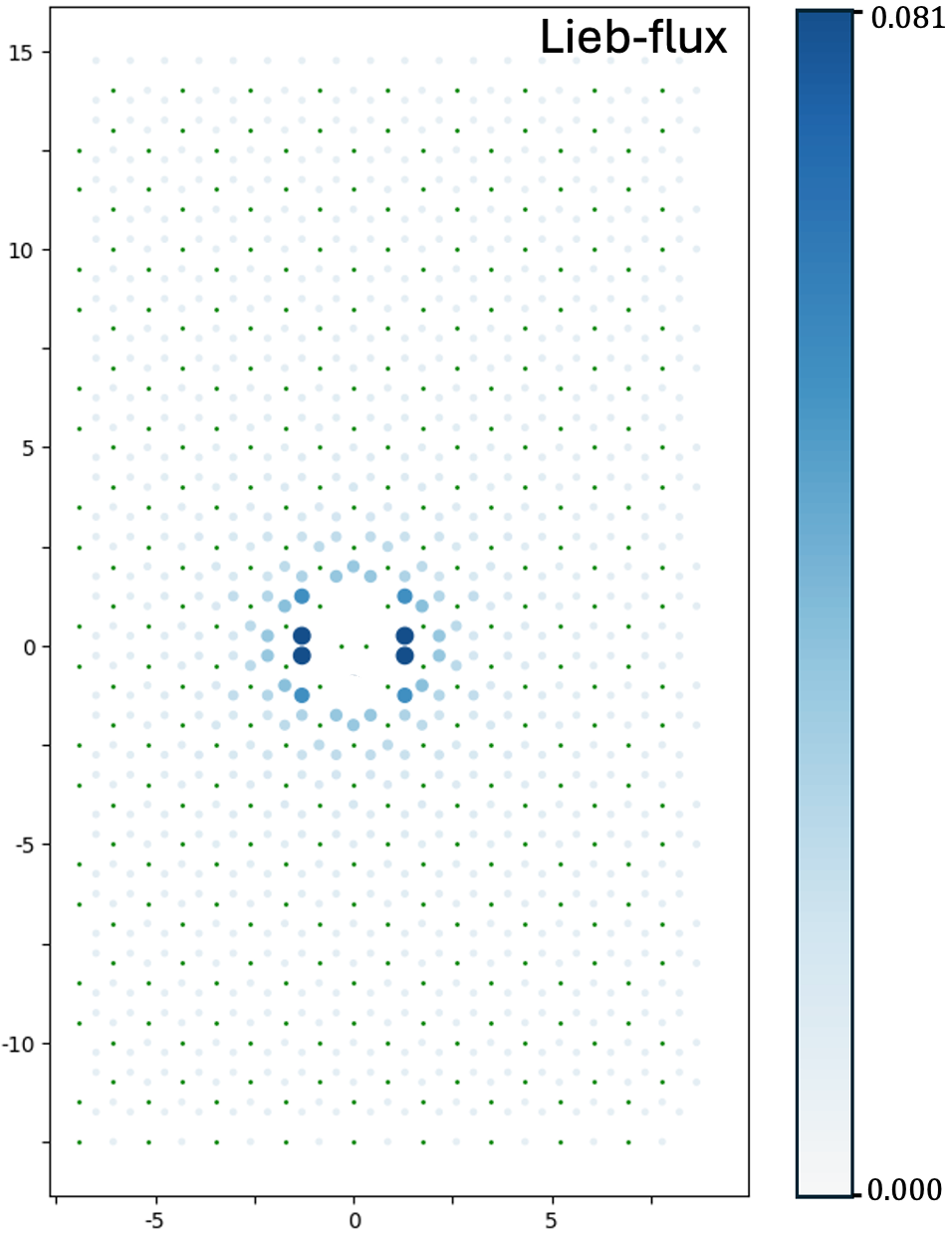}}
    \subfigure[]{
    \includegraphics[width=0.31\linewidth]{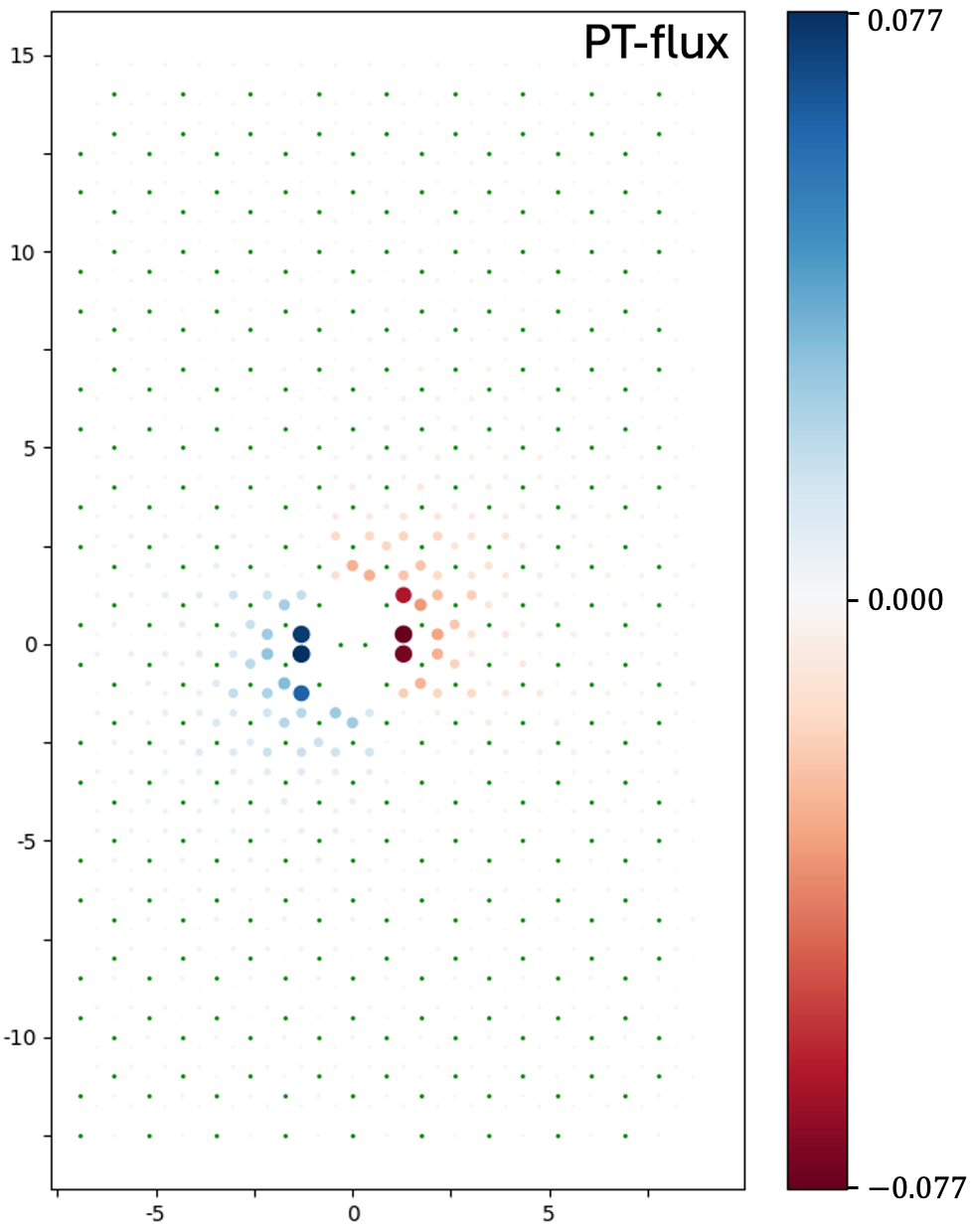}}
    \caption{{\bf Scalar spin chirality (SSC) induced by SW defects.}  (a) An array of SW defects with Lieb-flux (as shown on top panel) induces a net SSC for the quantum fluctuating spin-1/2, with average SSC approximately linear in the number of defects for typical defect densities (bottom panel, also see Fig. 2c of Ref.~\cite{seth_chiral_2025}). 
    Note the log-linear scale. (b,c) 2D distributions of SSC contributions for a single defect with (b) Lieb-flux, (c)  PT-flux. 
    Sites of the honeycomb lattice, modified by the SW defect, are shown as green dots. 
    The effective local SSC operator (Eq.~\ref{eq_ssc_majorana}) lives on second neighbor bonds (NNN) of this graph. SSC expectation values are shown as colored disks at NNN bond midpoints.   
    The SSC distribution is peaked near the defect but remains nonzero away from it. 
    Inner defect core SSC omitted for clarity.
    For (a) SSC remains always positive in the bulk, while for (b) it remains symmetrically distributed around zero (max and min values marked on color bar). These two profiles are consistent with the monopole and dipole distributions analogously to the local marker.}
    \label{fig_ssc 2d}
\end{figure*}

\subsection{Local Chern marker}

Here we describe how we compute the local Chern marker of the Majorana fermions. This is proportional to the topological orbital magnetization of the Majorana fermions. When charge fluctuations are considered this fermion magnetization also gives rise to  the  electron orbital magnetization. This can be seen  by the similar arguments for scalar spin chirality below \cite{shindou_orbital_2001, motrunich_orbital_2006, bulaevskii_electronic_2008}  or by its direct coupling to the Majorana fermion thermal Hall effect which is already known to couple to magnetic fields in the Kitaev model ~\cite{borhani_realspace_2025}. 

Mathematically, the local Chern marker $M(r)$ is a real space decomposition of the Chern number $C$ as $C =\frac{1}{N}\sum_{r\in \text{Bulk}}  M(r)$ with $N$ being the number of bulk sites. However, such decomposition is not unique and should be considered as the different lattice scale  manifestation of the coarse-grained underlying orbital magnetization created by the edge states. There exist multiple formulations of the local marker in the literature \cite{bianco_mapping_2011,dornellas_quantized_2022}.
The difference between the competing formulations becomes important in our case, where in the presence of an isolated defect the system remains nearly gapless  with no clear distinction between the bulk and the boundary. 

Here we use a variant of the local marker formulation based on Ref.~\cite{bianco_mapping_2011}. They consider the local marker 
$M_{0}(r)=4\pi \text{Im}\langle r|PxPyP|r\rangle$
where $P$ is the projection onto the occupied states. This can be further decomposed into $M_{0}(r)=\sum_R C(r,R)$ where $C(r,R)$ is a \textit{contribution map}, defining the contribution of site $R$ to local Chern marker at $r$. These are given by the following expression, with details in Ref.~\cite{borhani_realspace_2025}, and where $\theta_{x,R}$ is the step function along $x$ with a step up as $x$ crosses $R$ (i.e.\ it is 1 to the right of $R$ and 0 to its left) and similarly for $y$: 
\begin{align}
C(r, R) = 4\pi \text{Im}\langle r|P \theta_{x;R}P \theta_{y;R} P|r\rangle .
\end{align}

In a gapped system with a large and relatively homogeneous gap, defining a short correlation length, the bulk and boundary can be clearly distinguished in systems much larger than this correlation length. For the marker contribution, the function $C(r,R)$ is sufficiently local and only the lattice points at $R$ in the proximity of  $r$ contribute. However, in the present case of local mass terms arising from defects embedded in an extended nearly gapless background, $C(r,R)$ is a fairly non-local function with a long correlation length. This becomes problematic due to the nature of the local marker: the positions $R$ near an open boundary sample  contribute with an opposite sign due to the edge mode. In particular, for a large homogeneous gapped system, the total  local marker integrated over the bulk is exactly equal and opposite to the total local marker integrated over a region near the boundary. The boundary must always contribute a large opposite sign. In the present case, as we are interested in bulk contributions, these artificial boundary contributions must be avoided to get the physical answer. 

To minimize this issue, we extend $M_{0}$ by defining another coarse-grained marker $M_1$ in the following way, which also enables a rewriting in terms of
$C(r, R)$ constraining the sum over $R$ to a bulk region, as follows,
\begin{align}
    M_1(r) = 4\pi \text{Im}\langle r|P x_b P y_b P|r\rangle =\sum_{R\in\text{bulk}} C(r, R)
\end{align}
where $x_b = L_b^{-1}\sum_{R\in \text{bulk}} \theta_{x;R}$ and similarly for $y_b$. The operators $x_b$, $y_b$ are position operators $x$, $y$ whose eigenvalues saturate when approaching the boundaries. The parameter $b$ denotes when the saturation occurs, equivalently the width of the boundary strips that define the bulk region of the $R$ sum. Note this $b$ also determines the normalization $L_b^{-1}$ of the $\sum_R$ sum with $L_b$ being the linear width of the bulk region.  

This marker $M_1$ thus generalizes the previously used $M_0$  (which has $b=0$) and allows the marker to remain well defined at intermediate distances to the boundary. Deep in the bulk the marker is anyway $b$-independent.
Empirically we find $b=5$ is appropriate for the system sizes considered here, allowing $M(r)$ to avoid spurious boundary contributions even at the edges of the real space windows we choose to plot. 

Regarding the two equivalent formulations in terms of $x_b$ or $\theta$, we note that the expression in tersm of $x_b$, $y_b$ terms is computationally quicker. However the expression decomposed into the $C(r, R)$ sum may be more physically transparent for some purposes.

\begin{figure*}[t]
    \centering
    \subfigure[]{
    \includegraphics[width=0.35\linewidth]{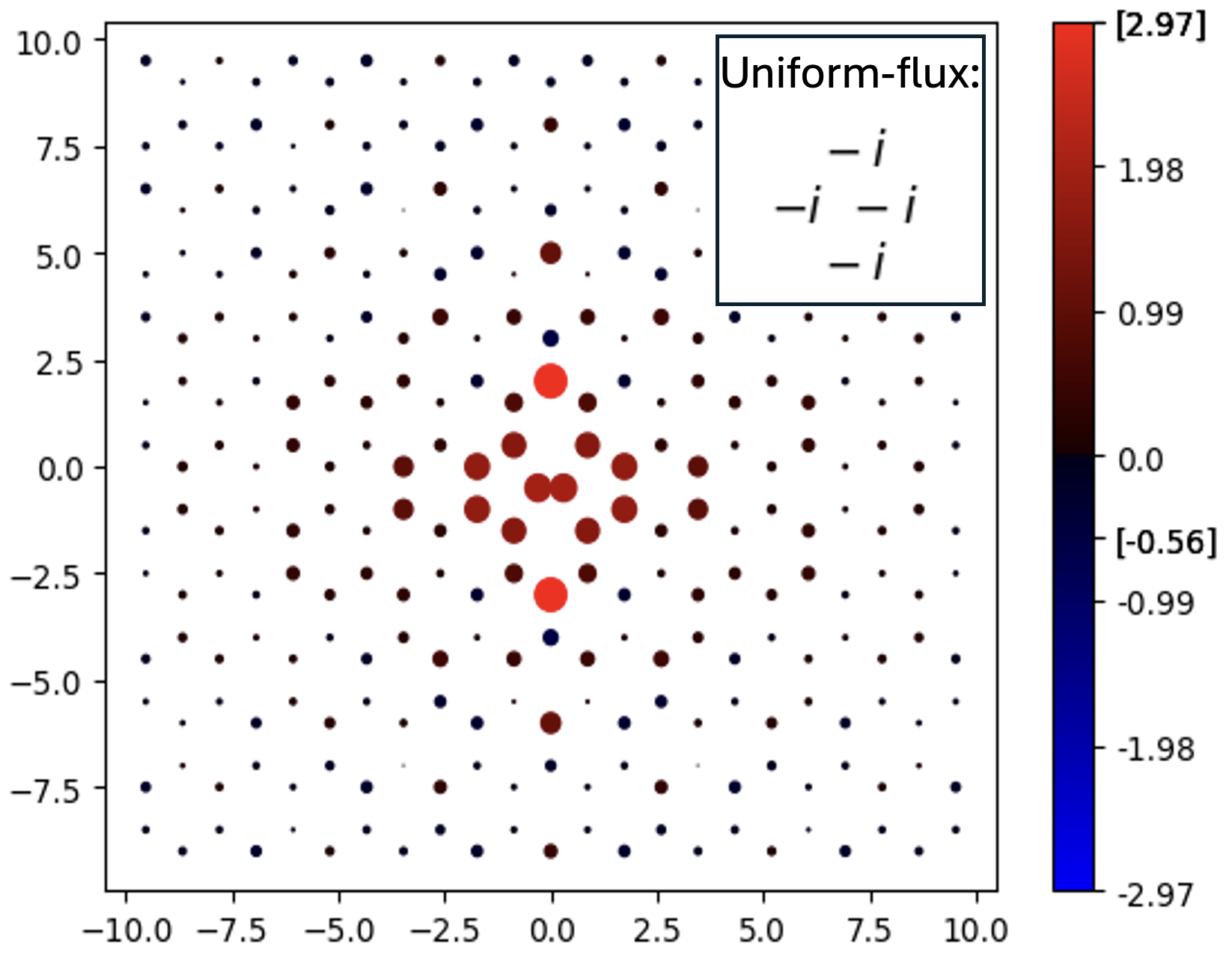}}
    \subfigure[]{
    \includegraphics[width=0.28\linewidth]{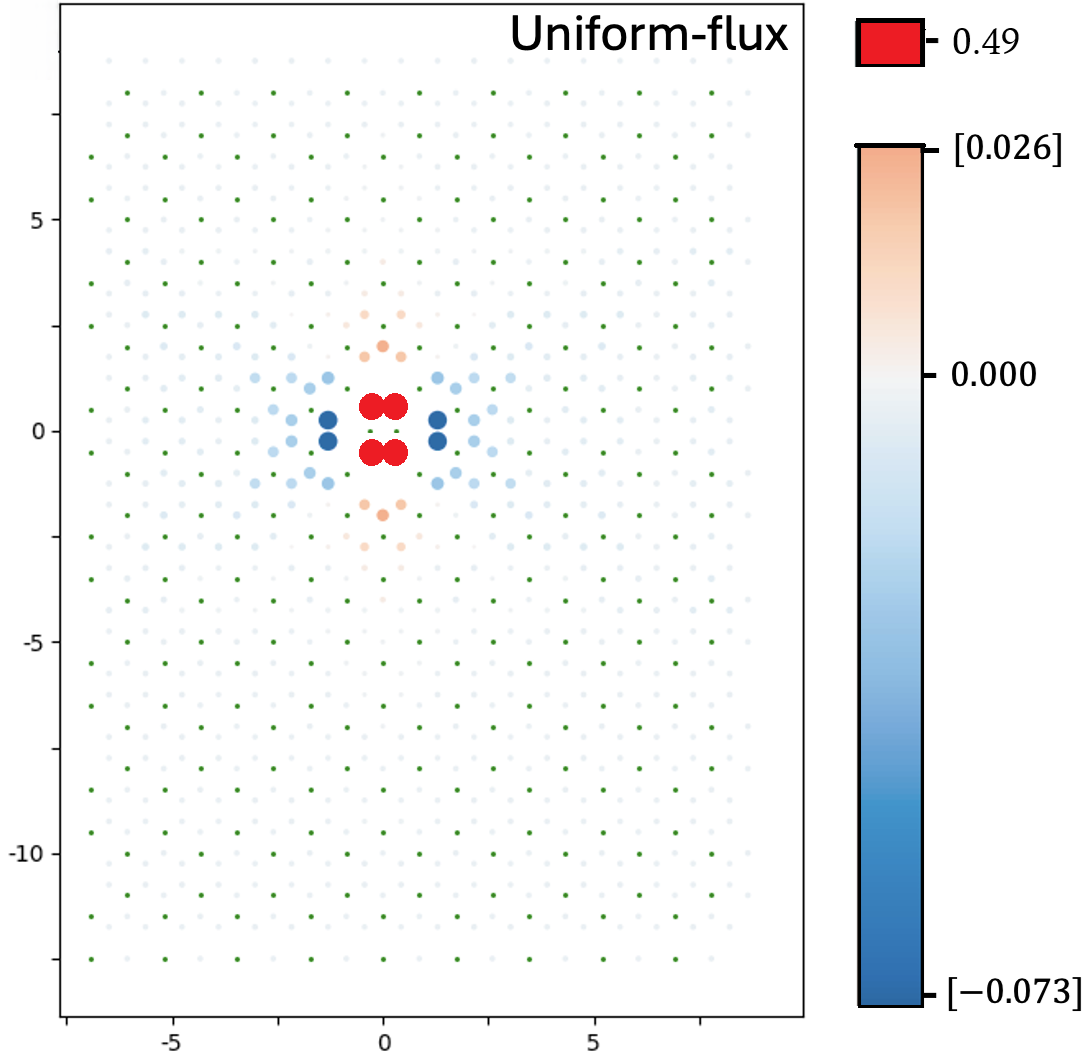}}
    \subfigure[ \label{fig_uniform_gap_chern}]{\includegraphics[width=0.34\linewidth]{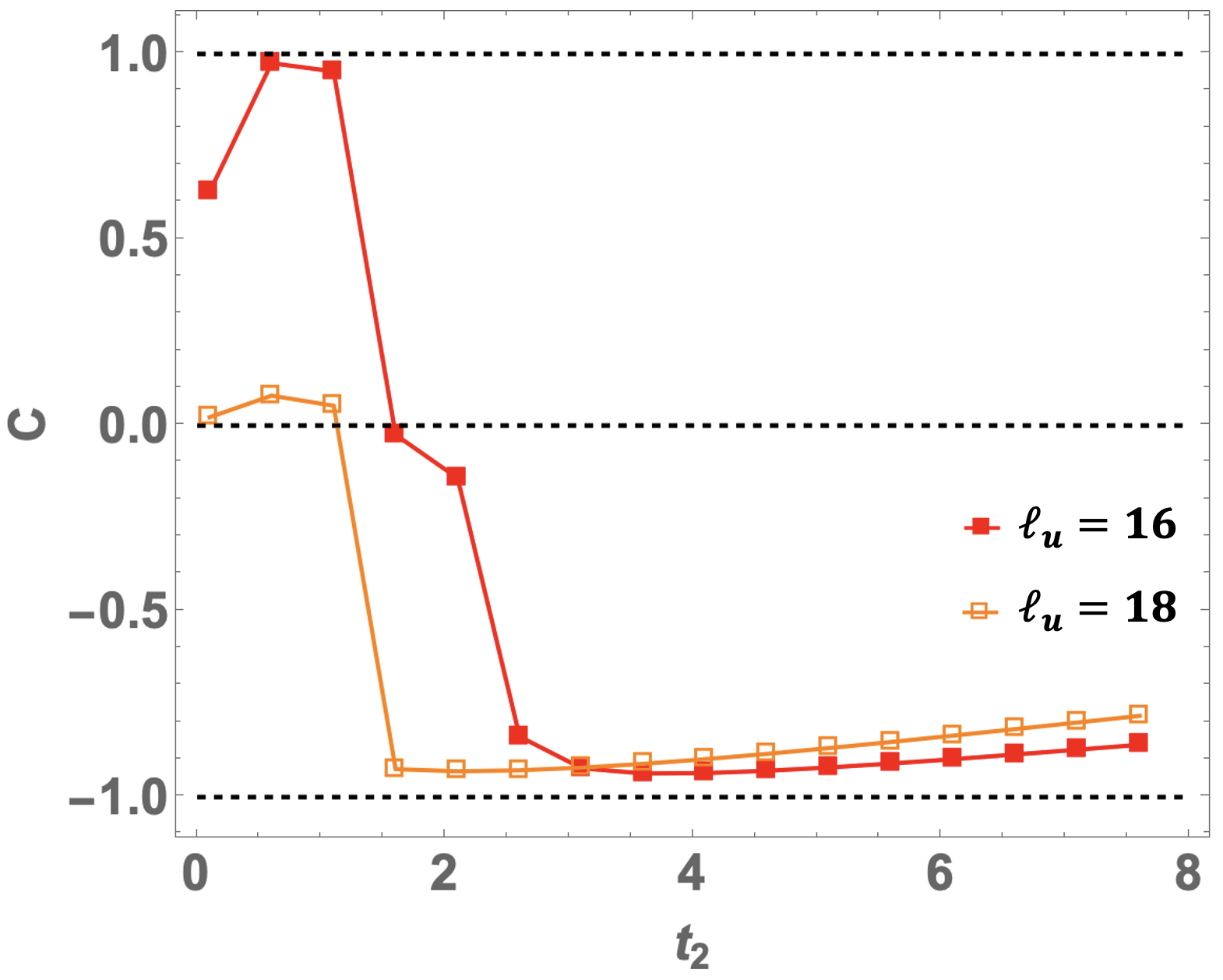}}
    \caption{{\bf Chirality induced by the Uniform-fluxes.}  
    (a) local marker and (b) scalar spin chirality (SSC) distribution induced by a single SW defect with Uniform-flux $\mu^z_U=+1$. The SW defect is here unperturbed  ($t_1=t_1'=t_2=1$).
    Plotting conventions are as in Figs. \ref{fig_BiancoResta_SW}, \ref{fig_ssc 2d}. 
    Both measures show a combination of a monopole and a quadrupole patterns of chirality, with dominant positive monopole term.
    In local marker, the quadrupole term is seen in the anisotropy of the largest contributions and in the presence of negative contributions (with largest amplitudes marked on color bar). In SSC the quadrupole is seen visually and the monopole is seen in innermost defect core bonds shown with an offset colorbar. 
    (c) 
    Density $n_d$ and perturbation $t_2$ dependence of Chern number for Uniform-flux defect superlattices. 
    A SW superlattice with defect density $n_d$ is obtained by placing a single defect in a unit cell of linear size  $\ell_u=n_d^{-1/2}$, and  the Chern number is computed with a 36$\times$36 integration mesh. $C$ is an integer; deviations of the computed $C$ from integers are artifacts implying small gaps. All defects carry the same Uniform-flux state, $\mu^z_U=+1$.%
    Chirality changes as a function of $t_2$ ($t_1=t_1'=1$) in a manner that is highly sensitive to defect densities.
    }
    \label{fig_uniform}
\end{figure*}

\subsection{ \label{sec_ssc_contruction}Scalar spin chirality}

Scalar spin chirality (SSC) is another real space measure of the chirality. It is locally defined as the triple product of three spin operators:
\begin{align}
    \hat{\chi}_{ijk}=\sigma_i\cdot \left(\sigma_j\times \sigma_k\right)
    \label{eq_ssc}
\end{align}
where $i,~j,~k$ are three consecutive sites on the 3-colored Kitaev graph, listed in counterclockwise order around the triangle plaquette they form. The scalar chirality is associated with the normal vector to this triangle plaquette. The orientation of the normal vector is set by the ordering of the sites as usual, with counterclockwise ordering corresponding to a vector oriented ``up" (right hand rule).
Note that the scalar spin chirality is odd under time-reversal, hence has zero expectation value in any flux sector of clean Kitaev model. But when a time-reversal breaking SW defect is introduced, it gives non-zero expectation value in the ground state flux sector.

SSC does not commute with the Kitaev spin Hamiltonian, hence does not necessarily conserve the flux. However the appropriate combination of spin components can potentially conserve the flux. When we compute the SSC expectation value in a fixed flux sector (such as the ground state), only the flux conserving terms remain. Projecting these terms into the conserved flux ground state sector gives rise to a Majorana hopping term between site $i$ and $k$ with strength $u_{ij}u_{jk}$:
\begin{align}
    \langle\hat{\chi}_{ijk}\rangle=i \langle  u_{ij} u_{jk} c_ic_k\rangle
    \label{eq_ssc_majorana}
\end{align}
This operator is a kinetic term in terms of the Majorana fermions between the second neighbor $i$ and $k$ sites.  Away from the defect, this term is always a non-bipartite term connecting two identical sublattices AA or BB. In the vicinity of the defect (odd sided plaquettes), the graph is not bipartite but the second neighbor bonds can still be defined.  

In terms of symmetry, the SSC is a TR breaking quantity associated with the normal vector to the triangle formed by the three sites. In other words, it has the symmetry of orbital magnetization, just like the topological orbital magnetization measured by the local marker. 
Charge fluctuations across the Mott insulating gap directly relate the SSC to the electron orbital magnetization along this axis. \cite{shindou_orbital_2001, motrunich_orbital_2006, bulaevskii_electronic_2008}

\subsection{Numerical results for Lieb-flux and PT-flux local marker}

The  local marker for Lieb and PT-fluxes are presented in Fig.~\ref{fig_BiancoResta_SW}. Due to the extended nearly gapless background, the local marker is not quantized, and has a sparse distribution. The ``\textit{Chern charge}" formulations~\cite{borhani_realspace_2025} suggests that each constituent disclination contributes to a sign of the chirality determined by its Frank angle sign $F$ and emergent gauge field flux $W =\pm i$ through the expression $q_M = - i F W$.
In terms of these Chern charges, the Lieb-flux gives rise to a contribution similar to a monopole of Chern charges, as 5 and 7 has opposite Frank angle and opposite fluxes. This feature is visible in Fig.~\ref{fig_BiancoResta_SW}(a). The 2D marker distribution is always positive in the bulk and shows an almost isotropic profile away from the defect consistent with symmetry of a monopole. %

In terms of Chern charges, PT-flux forms a dipole, since it contains 5-7 pairs with opposite fluxes. The dipole form is also required from the PT symmetry. Its local marker profile in Fig.~\ref{fig_BiancoResta_SW}(b) indeed shows the  dipolar symmetry with the marker field being distributed symmetrically about zero. As shown in the figure, the maxima and minima of the marker have exactly same magnitudes but opposite sign. %

\subsection{Numerical results for Lieb-flux and PT-flux scalar spin chirality}

Fig.~\ref{fig_ssc 2d} shows the  distribution of SSC around the SW for Lieb and PT-flux configurations. 
The contribution is large near the SW and decays away from the defect. %
Similar to the local marker result, the SSC shows the symmetry of a monopole, and a dipole, for the Lieb and PT-flux states, respectively.
For Lieb states, SSC is contributed only with a single sign. For PT, it has both positive and negative contributions symmetrically distributed around zero. %

We further compute the integrated SSC for Lieb configuration and study its dependence on the defect density in Fig.~ \ref{fig_ssc 2d}(a). We consider an array of defects with  a single defect placed in the unit cell. 
We find that the SSC density is directly proportional to the density of SW for dilute limit of defect densities, $\approx 5.3 n_d$ (see Fig.~\ref{fig_ssc 2d}(a)). This implies that SSC can serve as an order parameter for the spontaneous time-reversal breaking chiral instability of Kitaev spin liquid. For larger defect densities, SSC shows non-linear behavior arising from  interaction between the defects.

\subsection{ Uniform-flux chirality contributions: spatial redistribution and nonlinear dependence on density and on defect perturbations}

For the Uniform-flux case, the Chern charge formula predicts that the sign of the charges forms a quadrupolar distribution with two 5 (7) contributing either positive (negative) or negative (positive). However, this does not determine the magnitude of the  positive and negative charges. Generically, 5 and 7 having no symmetry relation, their Chern charge magnitudes are expected to be different. Thus the quadrupole gains an additional monopole component. This picture is consistent with our computed local marker (Fig.~\ref{fig_uniform}(a)) and SSC (Fig.~\ref{fig_uniform}(b)), where the asymmetries between the maxima and minima of their distributions measure the monopolar component. 

This  monopolar component is also seen in our T-matrix description (Sec. \ref{sec_tmatrix}).
Note that away from the defect core, local marker and SSC has different relative weight between the monopole and quadrupole contribution. This is generically expected as the local marker is sensitive to the nearly gapless background, whereas  SSC is sensitive to the local energy profile of the system.  
Nevertheless both SSC and local marker succeed in capturing both the quadrupole and monopole components. 
For finite defect densities (for example in array), the Chern charge magnitudes of the constituent disclinations are expected to get renormalized due to the complex interaction effects between the defects. It can generically lead to the possibility of flip or extinction of the chirality which we indeed find in the array case discussed above (Fig.~\ref{fig_majorana gap_chern}).

The
T-matrix result further predicts  that the monopole chirality flips with increasing $t_2$ perturbation. This again suggests a change of the magnitude of Chern charges with $t_2$. 
In Fig.~\ref{fig_uniform}(c), we compute the Chern number for array of SWs at various densities as a function of $t_2$. We find that  Chern number indeed changes with tuning $t_2$, and also it is quite sensitive to the defect density. %

\begin{figure*}[t]
    \centering    
    \includegraphics[width=2\columnwidth]{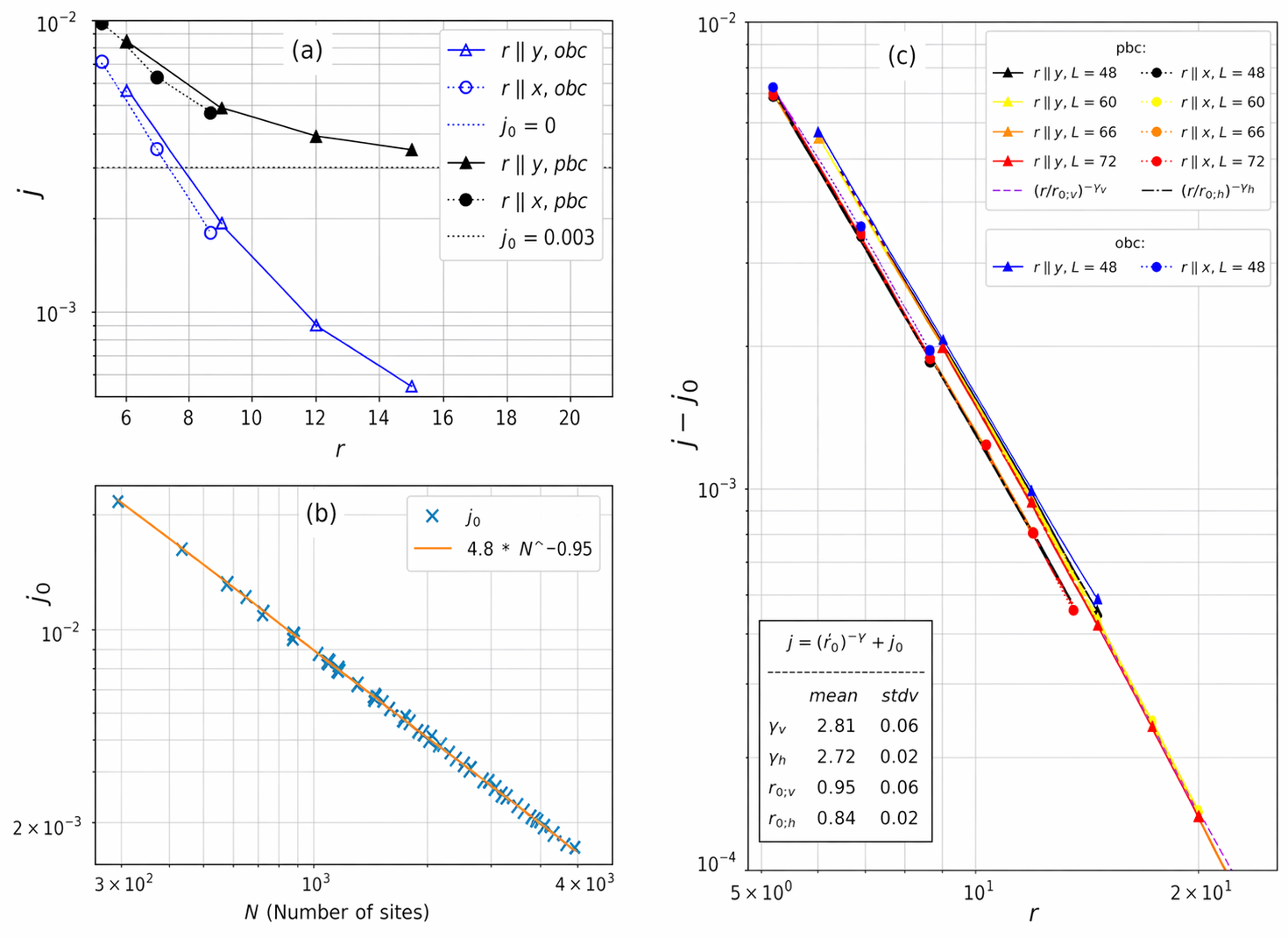}
    \caption{{\bf
    Emergent long range interaction between defect chiralities.}
    (a)  Energy difference $\Delta E$ between  aligned ($\mu^z_1=\mu^z_2$) and antialigned ($\mu^z_1=-\mu^z_2$) chirality configurations of two SW defects in a 48$\times$48 site system with open (OBC) (blue) or periodic (PBC) (black) boundary   conditions.
    The energy difference is plotted after dividing by 2, as $j\equiv \Delta E/2$, to suggest an interpretation in terms of effective Ising spin interaction of Eq.~\ref{eq_sw_int}.
    This numerically computed energy difference $j(r)$  vanishes  with increasing separation in OBC, but saturates to a non-zero value $j_0$ in PBC. 
    (b) The saturation offset $j_0$ is computed for various OBC system sizes and aspect ratios by placing two defects at maximal separation (corner and center) and computing  $\Delta E$.  Horizontal (along $x$) system size is taken to be divisible by 6 such that the Dirac cone maps to the $\Gamma-$point. The computed  $j_0$ decreases with number of sites $N$ as approximately $1/N$ with a slightly shifted exponent $N^{-0.95}$. 
    (c) After subtracting this offset $j_0$ the PBC $j-j_0$ and  the OBC $j$  energy differences agree, following the same power law $J(r)$ as a function of defect separation $|\vec{r}|$, with small but distinct differences between horizontal ($||x$, circles) and vertical ($||y$, triangles) separation vectors $\vec{r}$. 
    }
    \label{fig_gamma_pair int}
\end{figure*}

\section{\label{sec_long-range}Emergent long range interaction between defect chiralities}

We now restrict our consideration to the ground state Lieb-flux states, and consider the energetics of the sign of the Lieb-flux, $\mu^z$, when more than one defect is present in the system. We begin with two defects and then proceed to finite defect densities. 

In the following sections, we use two other methods to study the chirality interactions which further confirm our results reported in the companion paper (also described here in Eq.~\ref{eq_sw_int}).

\subsection{Effective two-defect chirality interactions in finite systems with PBC or OBC}

We place two SW defects with Lieb fluxes at  a separation vector $r$, measured between the inversion centers of the two defects.  Real space measurements of $r$ are in units of the honeycomb nearest neighbor bond length.
We use a finite system which can be either in open (OBC) or periodic (PBC) boundary conditions.
We then compute the energy difference between the aligned ($\mu^z_1=\mu^z_{2}$)   and antialigned ($\mu^z_1=-\mu^z_{2}$) flux configurations. 
We consider two kinds of defect separation vectors $\vec{r}$: $||x$ (horizontal separation `$h$') or $||y$ (vertical separation `$v$'). A minimum separation of the defects is always maintained such that they do not touch ($r_{\rm min}=3\sqrt{3}$ for $||x$ and $r_{\rm min}=6$ for $||y$, since the real space dimensions of a rectangular 4 site cell are $\sqrt{3} \times 3$). (Note that $r=0$ would correspond to the two defects being exactly on top of each other.)
In all cases we find that the aligned configuration has lower energy, consistent with a ferromagnetic interaction of $\mu^z$.

\begin{figure}[t]
    \centering
    \includegraphics[width=0.95\linewidth]{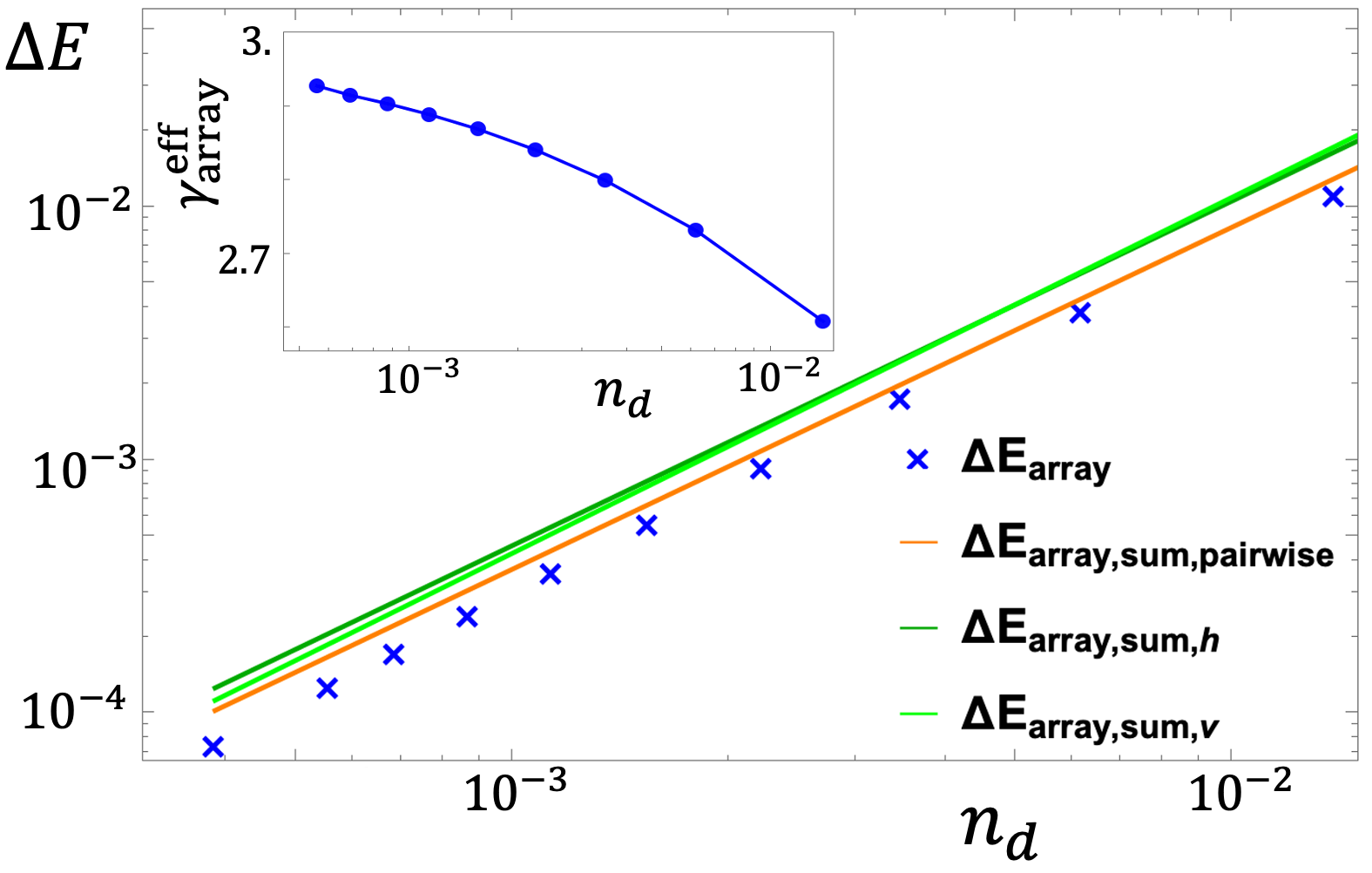}
    \caption{{\bf Stabilization energy of aligned chiralities for defect supercell array.} We consider superlattice arrays of Lieb-flux defects with  defect density $n_d$ by putting two defects in a supercell of linear size $\ell_u=(n_d/2)^{-1/2}$ (in corner and in center).  Main panel shows energy differences between ferromagnetic and N\'eel-antiferromagnetic  configurations of $\mu^z$, computed 
    numerically for SW defects (blue crosses) or by infinite sums $\sum_r$ (see Eq.~\ref{eq_energy_infinite sum}) of the approximated power law interaction  $J(r)=(r_0/r)^\gamma$ extracted from two defects (solid lines). To show the range of uncertainty in the power-law approximation, various parameter sets were used for $J(r)$: parameters for horizontal $h$ or vertical $v$ anisotropy from Fig.~\ref{fig_gamma_pair int} (dark and light green respectively), and parameters from the different extraction method of Ref.~\cite{seth_chiral_2025} Fig.~3a (orange). Up to an overall  prefactor of order 1--2, all the various methods agree. 
    {\bf Inset:} Effective power law exponent of $J(r)$ as a function of defect densities  computed from discrete derivative of blue crosses in main panel, showing the flow towards $\gamma_0=3$ at vanishing defect density. 
    The $\Delta E_\text{array}$,  $\Delta E_\text{array, sum, pairwise}$, and  $\gamma^{\rm eff}_\text{array}$  are also shown in Ref.~\cite{seth_chiral_2025} Fig.~3b; their explanation is here in Sec.~VII(B).    }
    \label{fig_power_law_array}
\end{figure}

In Fig.~\ref{fig_gamma_pair int}(a), we plot the energy difference for a small 48$\times$48 system with $N=2304$ sites. On this linear-log plot the $j(r)$ is not a straight lines showing a decay that is distinctly slower than exponential. 
While the energy difference decays to zero with increasing defect separation in the OBC case, it saturates to a finite nonzero value $j_0$ in the PBC case. 
In real space this can be interpreted as the effects of long range interactions wrapping around the torus to produce a finite energy shift. In momentum space this can be interpreted as directly sampling the Dirac cone gap generated by defect chiralities with finite size effects of the torus.

Computing $j_0$ for various systems sizes (Fig.~\ref{fig_gamma_pair int}(b)), we find that it is   only a function of the total number of sites $N$ regardless of torus aspect ratio. We find $j_0 \sim N^{-0.95}$ with an exponent whose magnitude is slightly suppressed from 1. We speculate that this is the result of a slow RG flow to an exponent of 1 in the $N\rightarrow \infty$ limit determined by the low energy Dirac theory, which is cut off by the finite defect density  and the associated nonlinear effects of multidefect scatterings.   
In any case
this $j_0$ term is well fitted by a power law and its finite size scaling is well controlled.
The relation between OBC and PBC pairwise interactions is thus well understood.

Adjusting for the energy shift $j_0$ of PBC, we find that PBC and OBC   both give  rise to  interactions that can be approximated as a power law $J(r)$ over the relevant range of densities and separations,  as shown by the straight line in log-log plot of Fig.~\ref{fig_gamma_pair int}(c), with a distinct anisotropy producing two different lines that depend on defect separation vector direction. 
The fit deviates from its linear behavior if we consider the defects being closer to the boundaries. To avoid this finite size effect, we always place the defects at the central region of the system and choose their maximum separation to be slightly less than 1/4 of the linear size of the system in the direction of the separation.

In Fig.~3a of the companion paper~\cite{seth_chiral_2025}, we used a different fitting method of the data for PBC, extracting the power law behavior with exponents similar to what we obtain here. The comparison is discussed in the next section, where we also compare to analytically summing the approximated power law form of the pairwise interactions across an infinite defect superlattice to compare with numerics we perform directly in this limit.

\subsection{Energy differences between ferromagnetic and antiferromagnetic defect arrays}
\label{sec:VIIB}

A complementary approach is to consider defects directly at finite density. To reliably extract an effective defect chirality interaction $J(r)$ requires the defect chirality system to be unfrustrated, leading us to consider defect superlattice arrays that are ``bipartite'' in the graph of defect-nearest-neighbors and hence admit unfrustrated N\'eel antiferromagnetic arrangements of SW defect chiralities $\mu^Z$. Thus we compute the energy difference between ferromagnetic and N\'eel antiferromagnetic arrangements of SW chiralities, allowing a comparison between the direct numerical computation and the analytical infinite sums of the approximated power law $J(r)$, as shown in Fig.~\ref{fig_power_law_array}.

An additional advantage of this comparison is that the consideration of finite defect density automatically includes any nonlinearities coming from interactions beyond two-body. We will show that the power law Ising model described by Eq.~\ref{eq_sw_int} still remains a  good approximation at this limit.

We consider an array of defects with two SWs placed per unit cell with linear dimension $\ell_u=(n_d/2)^{-1/2}$. The  defects are placed at the corner and at the center of the unit cell such that their  separation is approximately $\sqrt{3}\ell_u/2$. 
This defect superlattice can  be interpreted as a rectangular lattice with a two-site unit cell marked by the two defect positions, which can be considered as a sublattice and denoted by $\tilde{A}$ and $\tilde{B}$. 
Assigning $\mu^z_{\tilde{A}}=\mu^z_{\tilde{B}}$ (or alternatively, $\mu^z_{\tilde{A}}=-\mu^z_{\tilde{B}}$) fluxes on the defects generates a ferromagnetic (or alternatively, N\'eel antiferromagnetic) pattern in terms of these Ising variables. The energy difference between these two ordering (ferromagnetic and antiferromagnetic) is computed numerically (blue crosses in Fig.~\ref{fig_power_law_array}; also appears in Fig.~3 the companion paper~\cite{seth_chiral_2025}).

Using the approximated power law $J(r)$ interaction of the Ising chiralities, the energy difference can also be  approximately computed by performing  an infinite sum over pairwise interactions. This can be expressed as
\begin{align}  
    &\Delta E_\text{array, sum}
    =    \frac{1}{N}\sum_{i\in \tilde{A},\tilde{B}}\sum_{j\in\tilde{B}}2 J(r_{ij})
   =2 f_{\tilde{B}}(\gamma)  n_d^{\gamma/2}  
   \label{eq_energy_infinite sum}
\end{align}
where $N$ is the total number of lattice points and
\begin{align}
f_{\tilde{B}}(\gamma)=
\left(\frac{r_0}{\sqrt{2}}\right)^\gamma
    \sum_{\vec{n}\in \mathbb{Z}^2}
    \left|\vec{n u} + \vec{u}/2 \right|^{-\gamma}
    \label{eq_fb}
\end{align}
where $\vec{n u} \equiv  \left(n_x u_x, n_y u_y \right)$ with  $\vec{u}=(u_x,u_y)=(\sqrt{3},3)/2$. 
We explicitly compute 
$f_{\tilde{B}}(\gamma)$ 
by performing the infinite sum using $(r_{0;h},\gamma_h)$ and $(r_{0;v}, \gamma_v)$, and plot the corresponding  $\Delta E_\text{array,sum,h}$ and $\Delta E_\text{array,sum,v}$ in Fig.~\ref{fig_power_law_array} (dark and light green lines). %
The orange line in Fig.~\ref{fig_power_law_array} is obtained via performing the sum with $\gamma=2.7, ~r_0=0.75$ which were extracted using a different fitting method in the companion paper \cite{seth_chiral_2025}.

We are now in a position to summarize what we have learned about the emergent long range interactions between defect chiralities. First let us review. 
The extraction of the parameters for the long-range interaction is done here in Figures 8, and also using a complementary method in the companion paper \cite{seth_chiral_2025} Fig.~3a, using two defects in a finite system with open and with periodic boundary conditions. Here in Fig.~\ref{fig_power_law_array} we also consider two defects in a supercell producing a superlattice array.
We control the defect density by considering different systems sizes for the first two cases, and considering different supercell sizes for the later.

The comparison of pairwise interactions and $\Delta E$ energy differences of AFM and FM defect arrays is nontrivial since these do not have any simple relation. Within the power law approximated form of the interaction, an artificial relation can be constructed by analytically summing the power law. This is what we do in Fig~\ref{fig_power_law_array} main panel. The Fig~\ref{fig_power_law_array} main panel thus summarizes the differences between all the extraction methods: pairwise interactions extracted in two different methods and including both extremes of anisotropy (horizontal vs vertical separation), and separately, the numerically computed array energies. Remarkably, up to a factor of order 1--2, these all agree. 

The deviations of order 1--2 comes from the physically understood reasons of the expected anisotropy and the RG flow of $\gamma$ discussed below in Sec.~VII(C). These physically explainable uncertainties are much larger than any fit error bars. Though an uncertainty of order 1--2 is large for some purposes, for the present case, we are content with such an order 1 uncertainty in computing the observables such as the transition temperature $T_c$, since the simplified Kitaev model would not exactly describe real materials in any case, and the theoretical finding of a large $T_c$ of order $n_d$, described below, remains robust. 
Also robust is our conclusion that for all densities between $10^{-4}<n_d<10^{-2}$, the effective interaction  can be described as the chirality Ising model with long-range interaction, which over the relevant range of defect separations are well approximated by a power law $r^{-\gamma}$.

\subsection{Defect density dependence of interaction power law exponent $\gamma$}

To obtain the defect density dependent power law exponent, we compute discrete derivative  of the  data points (blue crosses in Fig.~\ref{fig_power_law_array}) obtained from the defect array consideration. This is plotted in the inset of Fig.~\ref{fig_power_law_array} and can be interpreted as an effective $\gamma^\text{eff}_\text{array}$. We expect with further decreasing defect density, $\gamma^\text{eff}_\text{array}$ flows to 3 which is the power law exponent of the RKKY interaction in Dirac cone systems~\cite{kogan_rkky_2011}. Substantial renormalization down from $\gamma_0=3$ is seen for the range of defect densities considered here. 
Note that the numerically computed  $\gamma^\text{eff}_\text{array}$  overestimates the effective $\gamma$ that would be appropriate for randomly spaced defects, since for $\gamma^\text{eff}_\text{array}$  all defects are equally spaced at maximal separation. 

This renormalization can be explained analytically in two steps. First, although  at finite defect density $n_d$ a small gap proportional to $n_d$ is opened at the Dirac cones,  it gives slow decay of the correlation functions compared to the typical defect separation of $n_d^{-1/2}$. Hence  long-range interaction between defects persist.
Second, a finite density of states is accumulated near the Dirac cone because of this small gap, effectively modifying the cone towards quadratic bands which are almost touching. For a perfect quadratic band touching, the RKKY interaction scales as $r^{-2}$~\cite{kogan_rkky_2011}. Thus the present case of partially formed quadratic band touching is expected to push $r^{-3}$ towards $r^{-2}$. This reduction of the exponent $\gamma$ can also be interpreted as an integral of the $r^{-2}$ local density of states of Dirac spinor $\sigma^z$ at each energy near the Dirac point (Dirac CFT has spin scaling dimension 2) \cite{liu_magnetic_2009}, but with this $k$ integral cut off at small $k$ by $n_d^{1/2}$, which again reduces $\gamma$ down towards 2.

The flow of an effective $\gamma$, and relatedly the uncertainty or error range in making a power law $J(r)$ approximation to the two-defect interaction form,  implies that though no exact power law describes this system,  the interactions are still slower than exponential and long ranged. This long ranged behavior will control the resulting physics of the defect chiralities.

\subsection{Results for anisotropic gapped Kitaev model}

In the gapped anisotropic Kitaev ``A" phase (here we consider $J_z>>J_x=J_y$), the interaction between SW chiralities can be computed analytically in a perturbative approach via computing the energy of the Wilson loop that encloses two odd-sided plaquettes from both the SWs.  The interaction can be shown to be ferromagnetic (see Appendix \ref{appen_gapped kitaev}). However, since in this case we integrate out the gapped bulk Majoranas,  the interaction becomes short-ranged, with the large distance asymptotic behavior  given by $\sim J_z e^{-\alpha r}$, where $\alpha\sim \frac{1}{{r'}_0} \log\left (\frac{J_z}{J}\right)$, and $r_0'$ is a length scale. %
A similar expression is known in the context of RKKY interaction in gapped graphene  \cite{kogan_rkky_2013}.

\section{\label{sec_tc}Instability to chiral spin liquid: $T_c$ for random defects and for defect superlattice arrays}

\subsection{Instability due to long range interactions}
At low temperatures, the long-range Ising interaction of defect chiralities aligns the SW fluxes into a ferromagnetic pattern leading to a spontaneous time-reversal broken non-Abelian chiral spin liquid phase. 
The chiral QSL is non-Abelian since the Majorana fermions have a nonzero Chern number $C=\pm 1$ for the case of flux alignment with $\mu^z=\pm 1$.

In the companion paper (Ref.~\cite{seth_chiral_2025}), we compute $T_{c,\text{random}}$ for the case of uncorrelated random defect locations, finding that $T_{c,\text{random}}$ scales linearly with defect density. This case is appropriate for localized crystallographic defects that form at random in a crystal. If defects impact their environment over a larger length scale, for example by a longer range elastic distortion, and if the defect density is relatively large, another possibility is that defects form at random positions but with a typical separation scale. In an idealized limit these spatially correlated defect positions can be considered to form a superlattice array. This array limit may also arise if defects are placed intentionally in a periodic fashion. We now proceed to compute $T_c$ in this limit, complementary to the uncorrelated-random defect limit. The comparison is plotted further below in Fig.~\ref{fig_phase diag_multidim}, which also includes the effects of defect perturbations $\sum_r \delta H_r$.

\begin{figure}[t]
    \centering
    \includegraphics[width=0.95\linewidth]{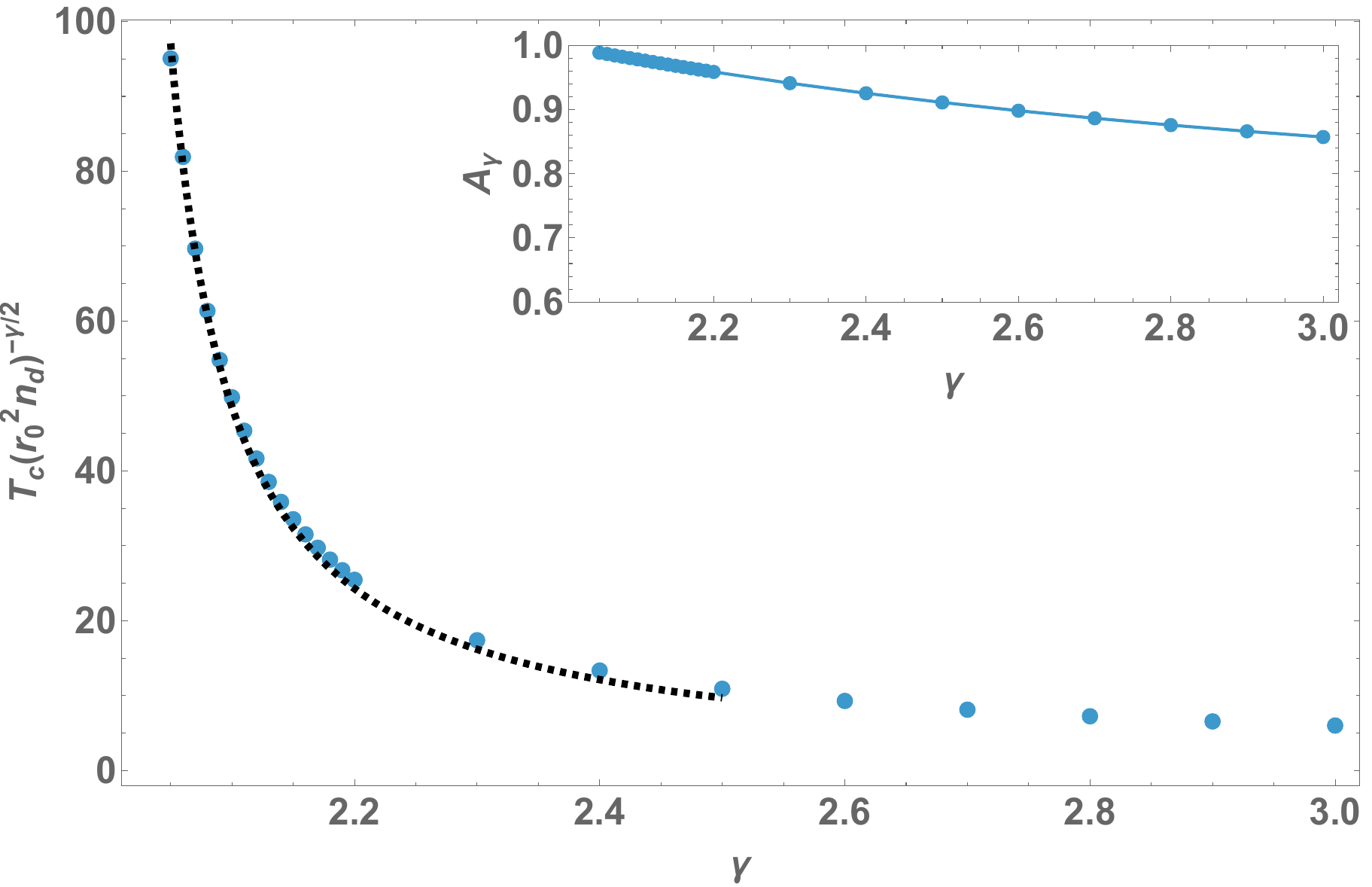}
    \caption{ {\bf Divergence of critical temperature with power law exponent $\gamma \rightarrow 2$.} 
    The critical temperature of the chiral QSL instability $T_c$, here computed for defect superlattice arrays,  is expressed in dimensionless form by dividing by $(r_0^\gamma n_d^{\gamma/2})$ and plotted as a function of $\gamma$. The dashed line show the fit for $T_c$ near $\gamma=2$ which diverges as $4.85/(\gamma-2)$. A similar divergence $\zeta(\gamma-1)$ or  $1/(\gamma-2)$ arises for random defects.  
    The inset shows the dimensionless  ratio $A_\gamma= T_c / \Delta E_\text{array}$ for infinite array sums, which remains an order 1 quantity with only weak $\gamma$ dependence even as $T_c$ diverges. 
    }
    \label{fig_tc_gamma}
\end{figure}

\subsection{Chiral spin liquid instability $T_c$ for superlattice defect array}

Due to the long range form of the interaction we compute the transition temperature of the Ising model in mean field theory. This implies an effective field on a single SW due to the rest of the SWs.
This field measures the energy gain due to the ferromagnetic configuration which  overcomes the entropic contribution to the free energy of the  flux disordered high temperature phase. 
This critical temperature is directly related to this mean field  
by $T_\text{c}=\frac{1}{N}\sum_{i,j}J(r_{ij})$ with $J(r) = (r/r_0)^{-\gamma}$ 
and is given by, 
\begin{align}
    T_\text{c, array, sum}=\frac{1}{N}\sum_{i,j}J(r_{ij})=\left(f_{\tilde{A}}(\gamma)+f_{\tilde{B}}(\gamma)\right)n_d^{\gamma/2}
    \label{eq_tc_sum}
\end{align}
where the sum over $i,j$ is implied to avoid $i=j$ and where $f_{\tilde{B}}(\gamma)$ is as given earlier in Eq.~\ref{eq_fb}, and 
\begin{align}
    f_{\tilde{A}}(\gamma)=
\left(\frac{r_0}{\sqrt{2}}\right)^\gamma
    \sum_{\vec{n}\in \mathbb{Z}^2 - \{\vec{0}\} }
    \left|{\vec{n u}} \right|^{-\gamma}
\end{align}
where 
again $\vec{n u} \equiv  \left(n_x u_x, n_y u_y \right)$ with  $\vec{u}=(u_x,u_y)=(\sqrt{3},3)/2$.

In Fig.~\ref{fig_tc_gamma}, we show $T_c$ as a function of $\gamma$ computed by performing the infinite sum. The result can be expressed in dimensionless form by dividing $T_c$ by $(r_0^\gamma n_d^{\gamma/2})$, which gives the quantity plotted in Fig.~\ref{fig_tc_gamma}, as a function of $\gamma$. 
Keeping $r_0$ fixed, we find that $T_c$ increases with  decreasing  $\gamma$. This suggests a possibility of tuning $T_c$ in  realistic materials with tuning defect densities; for example, as suggested by Fig.~\ref{fig_power_law_array} inset, higher defect densities can decrease an effective $\gamma$.

\begin{figure}[t]
    \centering
    \subfigure[]{\includegraphics[width=0.55\linewidth]{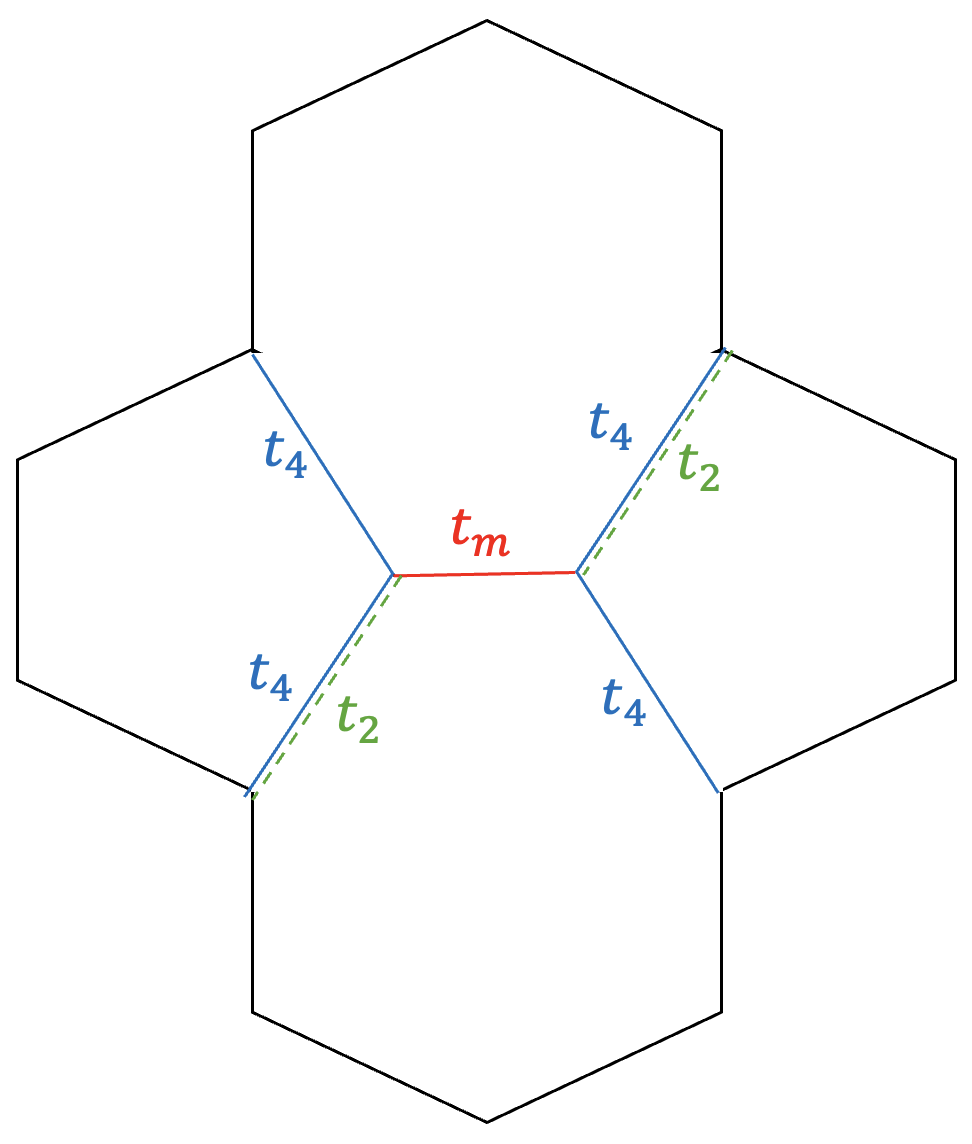}}
    \subfigure[]{\includegraphics[width=0.9\linewidth]{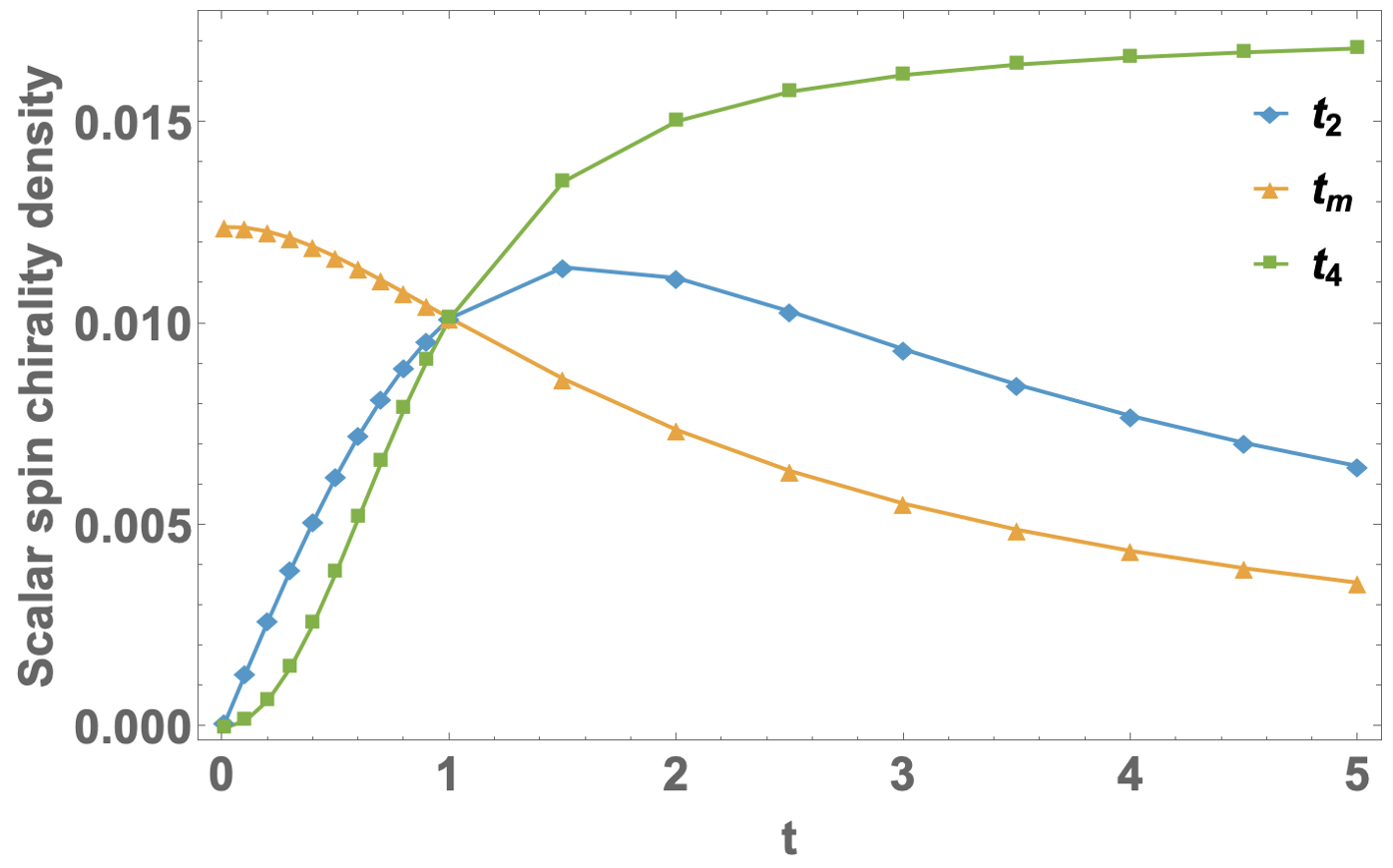}}
    \caption{{\bf Definition  of SW perturbation parameters and variation of SSC with  perturbations}: (a) Three types of perturbations are considered by modifying Kitaev exchanges on the defect core. The strength of the Kitaev exchange on the correspondingly colored bonds is scaled to take the value $J_K' = t J_K$, with $t\neq 1$ serving as a perturbation.  (b) Variations of average SSC density away from the defect core with $t_m, t_2$ and $t_4$ are computed for a periodic arrangement of  defects with a unit cell of  $36\times 36$ sites. The local defect perturbations produce substantial long range effects on the measurable SSC. }
    \label{fig_ssc_perturbation}
\end{figure}

\begin{figure*}
    \centering
\includegraphics[width=1\linewidth]{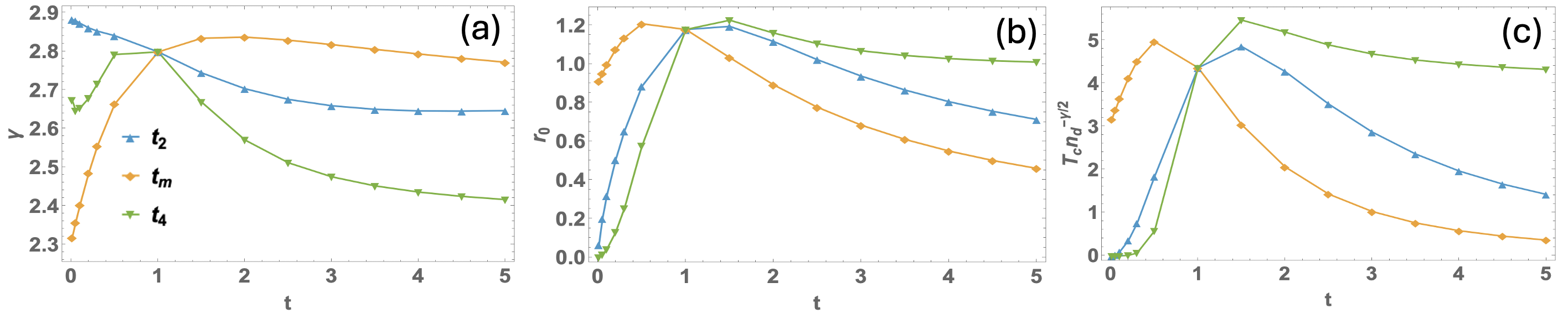}
    \caption{{\bf Variation of the long-range interaction parameters and $T_c$ 
    with various SW perturbations.} (a), (b) and (c)  show variation of $\gamma$, $r_0$ and the array $T_c$ with $n_d$ dependence factored out, respectively. The x-axis gives the value of the parameter $t$ corresponding to each symbol and color as shown in the legend, namely $t_2$ (blue up-triangles), $t_m$ (orange diamonds), and $t_4$ (green down-triangles). For each parameter, the Kitaev spin exchange  on the corresponding bonds is scaled by $t$ to take the value $J_K' = t J_K$. 
    }
    \label{fig_gamma_r0}
\end{figure*}

For $\gamma\rightarrow2$, $T_c$ diverges as  $1/(\gamma-2)$. This divergence  can also be seen if the infinite sum is replaced by a two-dimensional average-spaced sum in a continuum radial 1D approximation of the sums. In this approximation we replace the 2D sum $\sum_{\vec{n}}$ with a 1D radial sum over positive integers $n$ as $\sum_n 2 \pi n$ to get
\begin{align}
    T_\text{c, radial sum} &\propto\ r_0 ^\gamma n_d^{\gamma/2} \sum_n n^{-\gamma}\  2 \pi n 
    \nonumber \\
    &= 2\pi r_0 ^\gamma n_d^{\gamma/2}\ \zeta(\gamma-1)
\end{align}
with the Riemann zeta function $\zeta$. The well known divergence of $\zeta(s)$ at $s=1$ translates into a divergence of $T_c$ at $\gamma=2$ with the asymptotic form $1/(\gamma-2)$.

The ratio of the defect array mean field $T_c$ and the energy difference $\Delta E$ is a useful dimensionless quantity since it becomes independent of $n_d$ and of $r_0$, keeping only $\gamma$ dependence. We compute it through the infinite sums,
\begin{align}
    A_\gamma=\frac{T_\text{c, array, sum}}{\Delta E_\text{array, sum}}=\frac{f_{\tilde{A}}(\gamma)+f_{\tilde{B}}(\gamma)}{2f_{\tilde{B}}(\gamma)}
\end{align}
and plot it in the inset of Fig.~\ref{fig_tc_gamma} as a function of $\gamma$. It remains roughly 1 with only weak dependence on $\gamma$. Thus it serves as an useful conversion factor between $\Delta E_\text{array}$ and $T_c$ enabling $T_c$ to be estimated directly from the  numerical computations of energies of defects in an array, for example as given in Fig.~\ref{fig_power_law_array}.   

We further note that the effective Ising model derived here is based on a low energy assumption such that all the high energy flux excitations (other than local Lieb fluxes on SW) are inactive. Therefore, we expect the effective Ising interaction energy scale or equivalently $T_c$ to be cut off by the low lying SW PT-flux gap of $\approx 0.08 J_K$, which is independent of $n_d$.

\subsection{Short ranged interaction and ultralow-temperature instability for the anisotropic gapped Kitaev model}

For the case of the gapped Kitaev ``A" phase, the exponentially short ranged ferromagnetic  interaction  between the defect chiralities leads to a low temperature chiral spin liquid with an exponentially small critical temperature. In mean field it takes the value $\sim J_z e^{-\alpha/\sqrt{n_d}}$. For any realistic small $n_d$, this is much smaller than what we obtained for the gapless phase and likely too small to be observed.

 Though time reversal is broken, the Majorana bands are topologically trivial and have Chern number zero.
The resulting spin liquid in this case is Abelian, and hence is not a chiral spin liquid in the usual sense.

\section{\label{sec_perturbation}Impact of defect-induced perturbations within the Kitaev QSL}

In this section, we further study the effects of SW defects beyond the mere rearranged connectivity. 
As mentioned in Sec. \ref{sec_sw}, these effective perturbations can be of two types: perturbations that preserve the exact solvability of the Kitaev model and perturbations beyond  the Kitaev limit. Leaving the non-Kitaev type perturbations for future work, here we study the effect of the first type. Generically, such perturbations created by the defect  describes local inhomogeneities in the Kitaev exchanges. SW being a local defect, this effect is expected to be localized only in the surroundings of the defect. Here we model these inhomogeneities as an ultralocalized perturbations having support only on the SW core. 

We theoretically consider three kinds of modifications of the Kitaev exchanges, on three sets of bonds: (i) Middle bond of the SW, $t_m$; (ii) the four bonds connected to the middle bond, $t_4$; and (iii) two of these bonds connected to the middle bond, $t_2$, which are the nonbipartite bonds within the T-matrix formalism (see Fig.~\ref{fig_ssc_perturbation}(a)). 
These are dimensionless perturbation parameters which if multiplied with unperturbed Kitaev exchange $J_K$ gives the modified Kitaev exchanges on the respective bonds: $J_{m}=t_{m}J_K$, and similarly for other cases. 

Recall that we have already discussed some effects of defect perturbations for the case of $t_2$ in defining the T-matrix of the defects for various fluxes. Numerical results for the Uniform-flux defects, changing chirality with $t_2$ in a density-dependent manner, were shown in Fig.~\ref{fig_uniform}(c).
In this section we focus exclusively on the Lieb-flux states. 

We study these three types of perturbations separately, and  address how  they modify the long-range interaction parameters and the resulting $T_c$. We note that a realistic modeling of a SW defect in the Kitaev model is a complex problem and requires separate work. However, just looking at these three perturbations gives us interesting handle on tuning the low temperature non-Abelian chiral Kitaev physics, which can generate potential experimental insights.

\begin{figure*}
    \centering
    \includegraphics[width=1\linewidth]{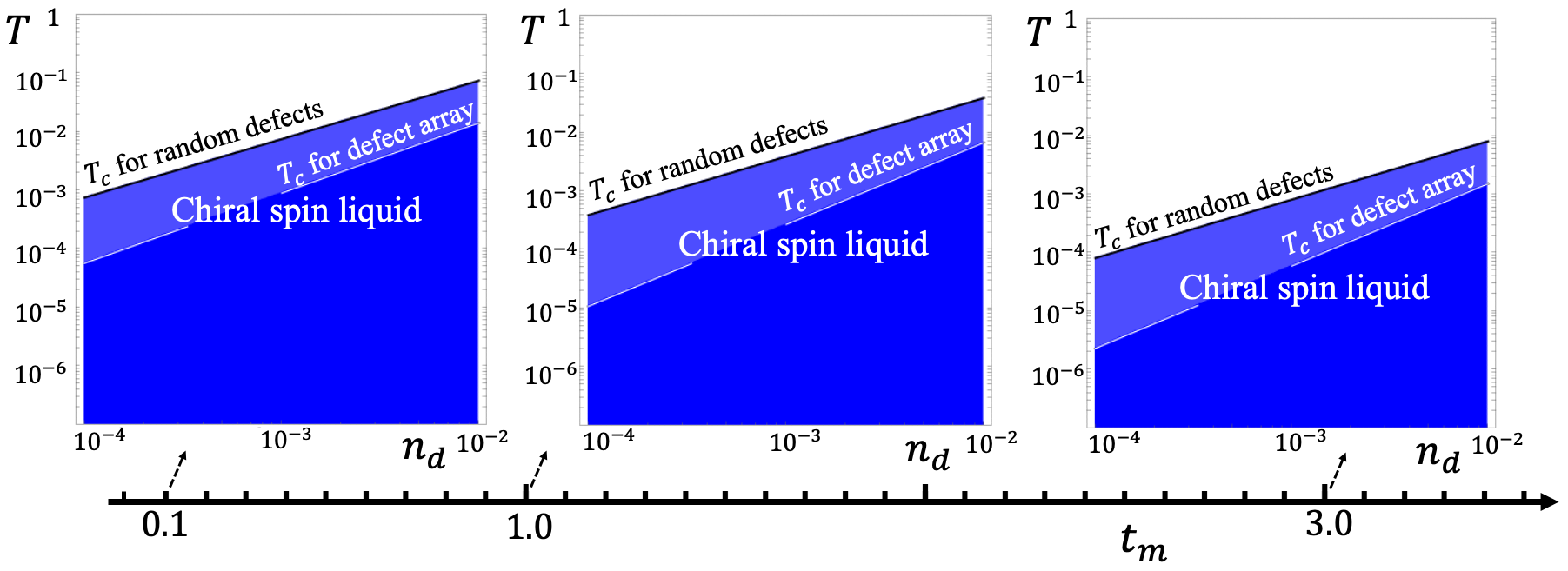}
    \caption{{\bf Chiral spin liquid instability  at low temperatures induced by the local SW defects  and dependence of $T_c$ on defect perturbation $t_m$.} Finite temperature $T$ phase diagrams as a function of defect density $n_d$ computed for several values of defect central bond strength $t_m=0.1, 1, 3$. Note $t_m=1$ are unperturbed SW defects; this phase diagram also appears in Ref.~\cite{seth_chiral_2025} Fig.~1. 
    The finite temperature transition $T_c$ into the chiral QSL is shown by the onset of colored region, with light blue marking $T_c$ for spatially-uncorrelated random defects and dark blue marking $T_c$ for spatially correlated defects in the superlattice array limit. }
    \label{fig_phase diag_multidim}
\end{figure*}

In Fig.~\ref{fig_ssc_perturbation}(b), we show the variation of the SSC of a single SW due to these perturbations. Recall that SSC is experimentally observable via any probe of time reversal breaking. This nontrivial variation already shows the interesting dependence of  time-reversal breaking order parameter on these perturbations. Note that SSC remains non-zero even at $t_m=0$ limit because SW then reduces to 5-12-5 defect, which retains time-reversal breaking effects due to the pentagons. On the contrary in both $t_2=0$ and $t_4=0$ limits, the defect reduces to only even sided plaquettes, hence produces zero SSC.

We further compute the variation of $\gamma$, $r_0$ and $T_c$  upon tuning these perturbations in Fig.~\ref{fig_gamma_r0}(a), (b) and (c), respectively. For $t_2\rightarrow0$ and $t_4\rightarrow0$, the defect reduces to a $T$ preserving one, hence the extent of chiral spin liquid phase vanishes with $T_c\rightarrow 0$. In contrast $t_m$ has larger effects  at small $t_m$ regime. Upon decreasing $t_m$,  $\gamma$ decreases towards 2, and a larger $T_c$ is expected. However, this $t_m$ perturbation alone also modifies $r_0$ which  impacts $T_c$. Due to their combined effect $T_c$ remains finite even for $t_m\rightarrow0$ limit (see Fig.~\ref{fig_gamma_r0}(c)). We  plot the variation of the finite temperature phase diagram in Fig.~\ref{fig_phase diag_multidim} for various representative values of $t_m$. Thanks to the reduction of $\gamma$ with decreasing $t_m$, we find that $T_c$ grows to $T_c \approx 10 n_d$ for small $t_m$. 

In future work it would be interesting to explore $t_m$ perturbations combined with other perturbations that may decrease $\gamma$ while also enhancing or preserving $r_0$, and thereby greatly enhance $T_c$. In any case these results already imply that defect perturbations can either suppress or amplify the instability to the chiral QSL.

\section{Discussion}

\label{sec:discussion}

In this work we considered  local SW defects which can be realizable in the Kitaev honeycomb spin liquid while approximately preserving  the edge sharing octahedra geometry and hence the  exact solvability of the model.
 The locality of the defect, which enables realizability of Kitaev bond labels, also allows us to perform a T-matrix analysis. This analysis indicates the local contributions to  chirality  generated by various defect flux configuration. 
 These contributions are nontrivial. For example, for the Uniform-flux states, changing the strength of core defect bonds can reverse  the chirality. 
We compute the chirality distributions  in real  space using the local Chern marker and scalar spin chirality, which lead to a distribution of electronic orbital magnetization.
The flux ground states are found to be exactly those states which have the largest net chirality contributions. 

By considering pairs of defects we identify an emergent interaction between the defect chiralities which turns out to be ferromagnetic and long-ranged. This interaction is consistent with energy differences extracted from finite defect densities. 
Its mechanism does not depend on the Majorana nature of the fermions, but rather only relies on the Dirac cone fermions coupled to a finite density of fluctuating Ising spins with each Ising spin producing a local mass term for the Dirac fermions. This unusual scenario of fluctuating random mass terms of Dirac fermions, where the mass terms are localized in fixed spatial positions but with dynamically fluctuating signs, may also be relevant for other systems. 

At finite densities and low temperatures, the resulting emergent interaction leads to ferromagnetic ordering of the defect chiralities via a finite temperature phase transition into a non-Abelian chiral spin liquid phase. 
The $T_c$ can be suppressed or enhanced by defect perturbations. 
This phase is well known from studies of the Kitaev QSL in a 111 magnetic field and is characterized by  quantized thermal Hall conductivity and associated chiral edge modes of Majorana fermions, producing various observables. Here it is also characterized by its scalar spin chirality and electronic orbital magnetization arising at zero external field, which can be probed with thermodynamic measurements or with local scanning probes. 
\cite{kitaev_anyons_2006,yokoi_halfinteger_2021,zhang_lowenergy_2025,zhang_probing_2025,koller_raman_2025, koller_spinlattice_2025,vasyukov_scanning_2013,finkler_selfaligned_2010,mclaughlin_local_2023,du_control_2017,grover_chern_2022}
In addition, the orbital magnetization produced here by crystallographic defects below $T_c$ offers unusually sharp finite-temperature signatures of $T=0$ fractionalization.

Let us now turn to open questions suggested by this study. 
The key questions concern energetic considerations for any particular material realization. Various perturbations are expected to arise locally near each defect. 
Here we studied only a subset of these perturbations, which involve local inhomogeneous modifications of the Kitaev bond strengths and thus preserve QSL solvability, thereby primarily affecting the chiral QSL $T_c$.
It would also be interesting to consider similar effects arising from 
a larger family of defect perturbations that preserve the QSL, to determine how large of a $T_c$ can be achieved.
Although the derived formulas for $T_c$ formally diverge with $\gamma\rightarrow2$, these rely on the effective long-range Ising model for defect chiralities  which is valid in the low-temperature regime with only Lieb-fluxes active. With other flux excitations, this long range interaction may be screened by other flux excitations, producing a cut-off to the diverging $T_c$ set by flux excitation energy scales. 
The extent to which such screening is effective, given that the Uniform-flux excitation also involves net chirality with presumably its own long range interaction, is an open question.

Most defect perturbations that do not preserve solvability are expected to suppress the QSL state. When these perturbations are weak they can be ignored because of the local mass terms (and local gap) generated by the defects. However, when they are strong, they can have substantial effects, at least locally and possibly globally. 
Relatedly, many Kitaev honeycomb candidate materials studied in the literature do not appear to have a true QSL ground state but have been studied for possible crossover effects at finite temperature. 
For the clean Kitaev QSL, which has a finite $T$ crossover scale associated with $\pi-$flux gap, such crossover behavior at finite $T$ can persist even for Hamiltonians whose ground state is not the QSL but which show proximate QSL nearby in the quantum phase diagram \cite{nasu_thermal_2015, rousochatzakis_quantum_2019}.

In the present case it would be interesting to consider two possibilities: either the full Hamiltonian parameters might belong to a non-QSL proximate phase, or defect perturbations might be strong enough such that they would destroy the QSL if the defects appeared at high density.
A general open question about defects in QSLs concerns the scenario when such strong defects appear at a low density. For example, there can be regions where the QSL is destroyed near defects even if the rest of the system preserves a percolating QSL state. In the present case additional new questions arise related to the chirality and the resulting time reversal breaking finite $T$ transition.  What happens to the emergent defect chirality, and to the transition, if the QSL is destroyed either globally or locally near defects? Do finite temperature crossover effects remain?
By explicitly studying various defect perturbations, including cases that enhance $T_c$ and cases that suppress the QSL itself, broader questions about finite-temperature behavior of QSLs or QSL-proximate phases may be addressed through the concrete setting of Eq.~\ref{eq_kitaev_1}.

\section*{Acknowledgments}

This work was supported by the U.S. Department of Energy, Office of Science, Basic Energy Sciences, under Early Career Award Number DE-SC0025478.
We thank Natalia Perkins,  Masaki Oshikawa and Gabor Halasz  for helpful discussions.

The data  that support the findings of this article are openly available at the GT Digital Repository \cite{kimchi_georgia_2026}.

\begin{appendix}

\section*{APPENDIX}

\section{\label{appen_re-im} Transformation of  Majorana Hamiltonian to complex fermion hopping}

The imaginary hopping Hamiltonian for the Majorana fermions  given in the Eq.~\ref{eq_majorana_im} of the  main text is equivalent to a complex fermion hopping model~\cite{borhani_realspace_2025} using the mapping $A_{i,j} c_i c_j\rightarrow A_{i,j} f^\dagger_i f_j + \text{h.c}$. This gives
\begin{align}
    H_{c}=-\frac{i}{2}\sum_{ i,j}  u_{ij}f^\dagger_if_j +{\rm h.c.}. 
\end{align}
Each bond is double counted in the sum. Note that the complex fermion representation doubles the number of degrees of freedom, but the model has same eigenspectra as the  Majorana fermions. Since $u_{ij}=\pm1$, $f-$fermion  model has imaginary hopping amplitudes on all the bonds. This can further be gauge transformed via $f_{r,B}\rightarrow i \bar{f}_{r,B}$ to another equivalent  Hamiltonian which has real hoppings on the  bonds connecting $A$ and $B$ sublattice (denoted below as \textit{bipartite} bonds).  Note that hopping on same sublattice bonds (\textit{non-bipartite} bonds) remain unchanged and has the imaginary hopping amplitudes. The real hopping Hamiltonian is then given by,
\begin{align}
    \bar{H}_{c}&=-\sum_{ \langle ij\rangle\in {\rm bipartite}}  u_{ij}\bar{f}^\dagger_i\bar{f}_j\nonumber\\
    &\hspace{2cm}-\sum_{ \langle ij\rangle\in \text{non-bipartite}}  i u_{ij}\bar{f}^\dagger_i\bar{f}_j+\text{h.c.}. 
\end{align}

Note that in $\langle ij\rangle$ notation  the bonds are counted only once. In the pristine Kitaev limit   on honeycomb lattice  there are no  non-bipartite terms. Ground state of the model is obtained by assigning zero fluxes on the hexagons, which we implement  via the gauge choice of $u_{ij}=1$. The resultant Hamiltonian is the same as well-known graphene and  is  equivalent  to Majorana Hamiltonian $H_0$  given in the main text. The  low energy theory of this model is obtained by approximating the Hamiltonian near the two Dirac cones, which is given by
\begin{align}
    P[\bar{H}_{0,c}] = -v_F\sum_{ q} {\bar{\psi}}^{\dagger}_{ q}\left(q_x\sigma^x\tau^z+q_y\sigma^y\right){\bar{\psi}}_{ q}.
    \label{eq_low energy_real}
\end{align}
where $\bar{\psi}_q=\left(\psi_{K+q,A},\psi_{K+q,B},\psi_{K'+q,A},\psi_{K'+q,B}\right)^T$. At low energy, the gauge transformation from the real hopping back to the imaginary hopping Hamiltonian is implemented by,
\begin{align}
    U=\text{exp}\left(i \frac{\pi}{4} \sigma^z\right)
\end{align}
which is  a $+\pi/2$ rotation around  $\sigma^z$.
Under this gauge transformation, the degrees of freedom transform as:
\begin{align}
    &\sigma^x\rightarrow\sigma^y~,~
    \sigma^y\rightarrow-\sigma^x,
    \label{eq_gauge}
\end{align}
and other Pauli matrices remain unchanged. This gauge transformation maps $P[\bar{H}_{0,c}]\rightarrow P[H_0]$ given in Eq.~\ref{eq_sw perturbation} of the main text.

\section{\label{appen_symmetry} Time-reversal (T) and inversion (P) symmetries  for real and imaginary hoppings }

\subsection{Imaginary hopping}

Inversion (P):
\begin{align}
    &f_{rA}\rightarrow -f_{-r,B} ~~~, ~~ f_{rB}\rightarrow f_{-r,A}\\
    &f_{kA}\rightarrow -f_{-k,B} ~~~, ~~ f_{kB}\rightarrow f_{-k,A}\\
    &\sigma^{x,z}\rightarrow-\sigma^{x,z} ~~,~~ \sigma^y\rightarrow\sigma^y\\
    &\tau^{x}\rightarrow\tau^{x}~~~,~~~ \tau^{y,z}\rightarrow-\tau^{y,z}
\end{align}

Time-reversal (T):
\begin{align}
    &f_{rA}\rightarrow f_{r,A} ~~~, ~~ f_{rB}\rightarrow -f_{r,B}\\
    &f_{kA}\rightarrow f_{-k,A} ~~~, ~~ f_{kB}\rightarrow -f_{-k,B}\\
    &\sigma^{x}\rightarrow -\sigma^{x} ~~,~~ \sigma^{y,z}\rightarrow\sigma^{y,z}\\
    &\tau^{x,y}\rightarrow\tau^{x,y}~~~,~~~ \tau^{z}\rightarrow-\tau^{z}
\end{align}

\subsection{Real hopping}

Inversion (P):
\begin{align}
    &\bar{f}_{rA}\rightarrow \bar{f}_{-r,B} ~~~, ~~ \bar{f}_{rB}\rightarrow \bar{f}_{-r,A}\\
    &\bar{f}_{kA}\rightarrow \bar{f}_{-k,B} ~~~, ~~ \bar{f}_{kB}\rightarrow \bar{f}_{-k,A}\\
    &\sigma^{x}\rightarrow\sigma^{x} ~~,~~ \sigma^{y,z}\rightarrow-\sigma^{y,z}\\
    &\tau^{x}\rightarrow\tau^{x}~~~,~~~ \tau^{y,z}\rightarrow-\tau^{y,z}
\end{align}

Time-reversal (T):
\begin{align}
    &\bar{f}_{rA}\rightarrow \bar{f}_{r,A} ~~~, ~~ \bar{f}_{rB}\rightarrow \bar{f}_{r,B}\\
    &\bar{f}_{kA}\rightarrow \bar{f}_{-k,A} ~~~, ~~ \bar{f}_{kB}\rightarrow \bar{f}_{-k,B}\\
    &\sigma^{x,z}\rightarrow \sigma^{x,z} ~~,~~ \sigma^{y}\rightarrow-\sigma^{y}\\
    &\tau^{x,y}\rightarrow\tau^{x,y}~~~,~~~ \tau^{z}\rightarrow-\tau^{z}
\end{align}

\section{\label{appen_tmatrix} Details of T-matrix computations}

In this section, we will derive the low energy effective T-matrix in different flux sectors considering the real hopping on the bipartite bonds of the Kitaev graph.
The T-matrix for a defect potential $V$  is given by,
\begin{align}
    T(E)=V (1-G_0(E) V)^{-1}
\end{align}
where $G_0(E)$ represent the Green's function of the unperturbed Hamiltonian at energy $E$. In the present case, the unperturbed  zero-flux clean Kitaev model Green's function can be obtained by  setting a uniform gauge choice of $u_{ij}=1$. This reduces the Majorana hopping problem to the hopping problem of graphene. Such low energy Green's function is already known~\cite{kot_band_2020} which we use here. We here give the details of the T-matrix analysis for different flux sectors, in the real hopping gauge.

For each flux state we now proceed to give: the defect impurity potential that generates the SW defect with corresponding flux; its projection to the low energy Dirac cones; and the resummed T-matrix result. The tilde over parameters reminds us that these results are written in the gauge of real hopping amplitudes. 

\begin{widetext}

\textbf{Lieb-flux:}
\begin{align}
    &\bar{V}^\text{SW}_\text{Lieb} = \left(\bar{f}^\dagger_{R,A} , \bar{f}^\dagger_{R,B}\right)\left(
t_1 \sigma^0 - i  t_2\mu^z \sigma^0
\right)\left(
    \begin{array}{c}
         \bar{f}_{R+d_+,A}  \\
         \bar{f}_{R-d_+,B} 
    \end{array}
    \right)+\text{h.c.}
 \\
    &P[\bar{V}_\text{Lieb}^\text{SW}]=\frac{1}{\mathcal{N}_c}\bar{\psi}^\dagger_q\left(\sqrt{3}t_2 \mu^z\tau^z\sigma^z-t_1    \vec{m}_t\cdot(\tau^x,\tau^y)\sigma^x-2t_1\vec{\phi}^*\cdot (\sigma^x,\sigma^y\tau^z)\right)\bar{\psi}_q\\
    & P[\bar{T}_\text{Lieb}^\text{SW}]=\frac{1}{\mathcal{N}_c}\bar{\psi}_q^\dagger \tilde{f}_1\left(\tilde{a}_1 t_2\mu^z\tau^z\sigma^z
    -\tilde{b}_1 \vec{m}_t\cdot\left(\tau^x,\tau^y\right)\sigma^x-\tilde{c}_1\sigma^x-\tilde{d}_1\sigma^y\tau^z\right)\bar{\psi}_q\\
    &  \tilde{f}_1= \frac{1}{(4\pi-3\sqrt{3})(t_1^2+t_2^2)+18\pi-12\pi t_1} ,~~ \tilde{a}_1= 18\sqrt{3}\pi ,~~ \tilde{b}_1=18\pi t_1-(9\sqrt{3}+6\pi)(t_1^2+t_2^2)\\
    &\tilde{c}_1= 18\pi t_1+(3\pi-9\sqrt{3})(t_1^2+t_2^2),~~ \tilde{d}_1= 9\sqrt{3}\pi(t_1^2+t_2^2-2t_1) 
\end{align}

\noindent \textbf{PT-flux:}
\begin{align}
    &\bar{V}^\text{SW}_\text{PT}= \left(\bar{f}^\dagger_{R,A} , \bar{f}^\dagger_{R,B}\right)\left(
t_1 \sigma^0 + i  t_2\mu_{PT}^z \sigma^z
\right)\left(
    \begin{array}{c}
         \bar{f}_{R+d_+,A}  \\
         \bar{f}_{R-d_+,B} 
    \end{array}
    \right)+\text{h.c.}
\\
    &P[\bar{V}_\text{PT}^\text{SW}]=\frac{1}{\mathcal{N}_c}\bar{\psi}^\dagger_q\left(\sqrt{3}\mu^z_\text{PT}t_2 \tau^z-t_1    \vec{m}_t\cdot(\tau^x,\tau^y)\sigma^x-2t_1\vec{\phi}^*\cdot (\sigma^x,\sigma^y\tau^z)\right)\bar{\psi}_q\\
    &P[\bar{T}_\text{PT}^\text{SW}]=\frac{1}{\mathcal{N}_c}\bar{\psi}_q^\dagger \tilde{f}_2 \left(\tilde{a}_2 t_2 \mu^z_\text{PT}\tau^z-\tilde{b}_2\vec{m}_t\cdot\left(\tau^x,\tau^y\right)\sigma^x-\tilde{c}_2\sigma^x-\tilde{d}_2\sigma^y\tau^z
    \right)\bar{\psi}_q\\
    & \tilde{f}_2=\frac{1}{(4\pi-3\sqrt{3})(t_1^2-t_2^2)+18\pi-12\pi t_1},~~    \tilde{a}_2= 18\sqrt{3}\pi,~~ \tilde{b}_2=18\pi t_1-(9\sqrt{3}+6\pi)(t_1^2-t_2^2),\\
    & \tilde{c}_2= 18\pi t_1-(9\sqrt{3}-3\pi)(t_1^2-t_2^2),~~ \tilde{d}_2= -9\sqrt{3}\pi(t_2^2-t_1^2+2t_1) 
\end{align}

\noindent \textbf{Uniform-flux:}
\begin{align}
    &\bar{V}^\text{Uniform}_\text{PT}= \left(\bar{f}^\dagger_{R,A} , \bar{f}^\dagger_{R,B}\right)\left(
t_1 \sigma^0 + i  t_2\mu_{PT}^z \sigma^z
\right)\left(
    \begin{array}{c}
         \bar{f}_{R+d_+,A}  \\
         \bar{f}_{R-d_+,B} 
    \end{array}
    \right)+2t_1 \bar{f}^\dagger_{R,B}\bar{f}_{R+d_-,A}+\text{h.c.}
\\
    &P[\bar{V}_\text{Uniform}^\text{SW}]=\frac{1}{\mathcal{N}_c}\bar{\psi}^\dagger_q\left(\sqrt{3}\mu^z_\text{U}t_2 \tau^z-t_1    \vec{m}_t\cdot(\tau^x,\tau^y)\sigma^x
    -t_1'    \vec{m}_t'\cdot(\tau^x,\tau^y)\sigma^x-2t_1\vec{\phi}^*\cdot (\sigma^x,\sigma^y\tau^z)-{t_1'}\vec{\phi}\cdot (\sigma^x,\sigma^y\tau^z)
    \right)\bar{\psi}_q\\
    &P[\bar{T}_\text{Uniform}^\text{SW}]=\frac{1}{\mathcal{N}_c}\bar{\psi}_q^\dagger \tilde{f}_3 \left(\tilde{a}_3 t_2 \mu^z_\text{U}\tau^z\sigma^z+\tilde{a}_3't_2 \mu^z_\text{U}\tau^z
    -\vec{m}_t\cdot\left(\tilde{b}_3\tau^x,\tilde{b}'_3\tau^y\right)\sigma^x-\tilde{c}_3\sigma^x-\tilde{d}_3\sigma^y\tau^z\right)\bar{\psi}_q\\
    &\tilde{f}_3=\left(2\pi \left(9-6 t_1'-6 t_1+2t_1 t_1'+2t_1^2-2t_2^2\right)+3\sqrt{3}(-t_1^2+2t_1{t'}_1+t_2^2)\right)^{-2}\\
    & \tilde{a}_3=18\sqrt{3}\pi t_1'\sqrt{\tilde{f}_3}\\
    & \tilde{a}_3'=-18\sqrt{3}\pi (t_1'-1)\sqrt{\tilde{f}_3}\\
    &\tilde{b}_3=-\left[6\pi^{2}\bigl(9 - 6 t_{1}' -6t_1 + 2 t_{1}^2 + 2t_1t_{1}' - 2 t_{2}^{2}\bigr)
     \bigl(- 6 t_{1}'    -6t_1 + 2 t_{1}^2 + 2t_1t_{1}'  - 2 t_{2}^{2} - 2 t_{2}^{2}\bigr)\right.
  \;\nonumber\\
  &\hspace{2cm}+\;\left.9\sqrt{3}\,\pi\,\bigl(18 - 6 t_{1}' 
      -6 t_1+ 2 t_{1}^2 + 2t_1t_{1}' 
      - 2 t_{2}^{2}\bigr)
     \bigl(t_{1}^{2} - 2t_{1} t_{1}' - t_{2}^{2}\bigr)
  -
  27\bigl(-t_{1}^2 +2 t_1t_{1}' + t_{2}^{2}\bigr)^{2}\right]\\
    &  \tilde{b}'_3=18\sqrt{3}\pi (t_1-1)t_1'\sqrt{f_3}\\
    &\tilde{c}_3=6 \pi^{2}\,\bigl(9 - 6 {t'}_{1} 
    -6 t_1+ 2 t_{1}^2 + 2t_1{t'}_{1} 
    - 2 t_{2}^{2}\bigr)\,
   \bigl(t_{1}^2+6t_1  - 8 t_1{t'}_{1} 
    + 6 {t'}_{1} - t_{2}^{2}\bigr)
\;\nonumber\\
&\hspace{2cm}+\;-9\sqrt{3}\,\pi\,\bigl(18 + 5 t_{1}^{2} - 6 {t'}_{1} 
    - 6t_1 - 4t_1 {t'}_{1} 
    - 5 t_{2}^{2}\bigr)\,
   \bigl(t_{1}^{2} - 2t_{1}{t'}_{1} - t_{2}^{2}\bigr)
\;+\;
81\bigl(-t_{1}^2 +2t_1 {t'}_{1} + t_{2}^{2}\bigr)^{2}\\
    & \tilde{d}_3=9\sqrt{3}\pi (t_1^2-t_2^2-2t_1+2t_1')\sqrt{f_3}
\end{align}
In the above expression $\vec{m}_t=({\rm Re},{\rm Im})e^{i(K-K')\cdot r}$, $\vec{m}'_t=({\rm Re},{\rm Im})e^{i(K-K')\cdot r+i\pi/3}$, $\vec{\phi}=({\rm Re},{\rm Im})e^{i\pi/3}$, and $\vec{\phi}^*$ is complex conjugate of $\vec{\phi}$. These T-matrix results can be gauge transformed to the imaginary hopping case by applying \ref{eq_gauge}. This gives Eq.~\ref{eq_tmatrix_lieb}-\ref{eq_tmatrix_uniform} in the main text with the coefficients
\begin{align}
    &a_1=\tilde{a}_1,~~a_2=\tilde{a}_2,~~a_3=\tilde{a}_3,~~a'_3=\tilde{a}_3, ~~b_1=\tilde{b}_1,~~b_2=\tilde{b}_2,~~b_3=\tilde{b}_3,~~b'_3=\tilde{b}_3',\nonumber\\
    &~~c_1=\tilde{c}_1,~~c_2=\tilde{c}_2,~~c_3=\tilde{c}_3,~~ d_1=-\tilde{d}_1,~~d_2=-\tilde{d}_2,~~d_3=-\tilde{d}_3.
\end{align}

\end{widetext}

\section{\label{appen_gapped kitaev} Flux interactions in the gapped phase}

\begin{figure}
    \centering
    \includegraphics[width=0.95\linewidth]{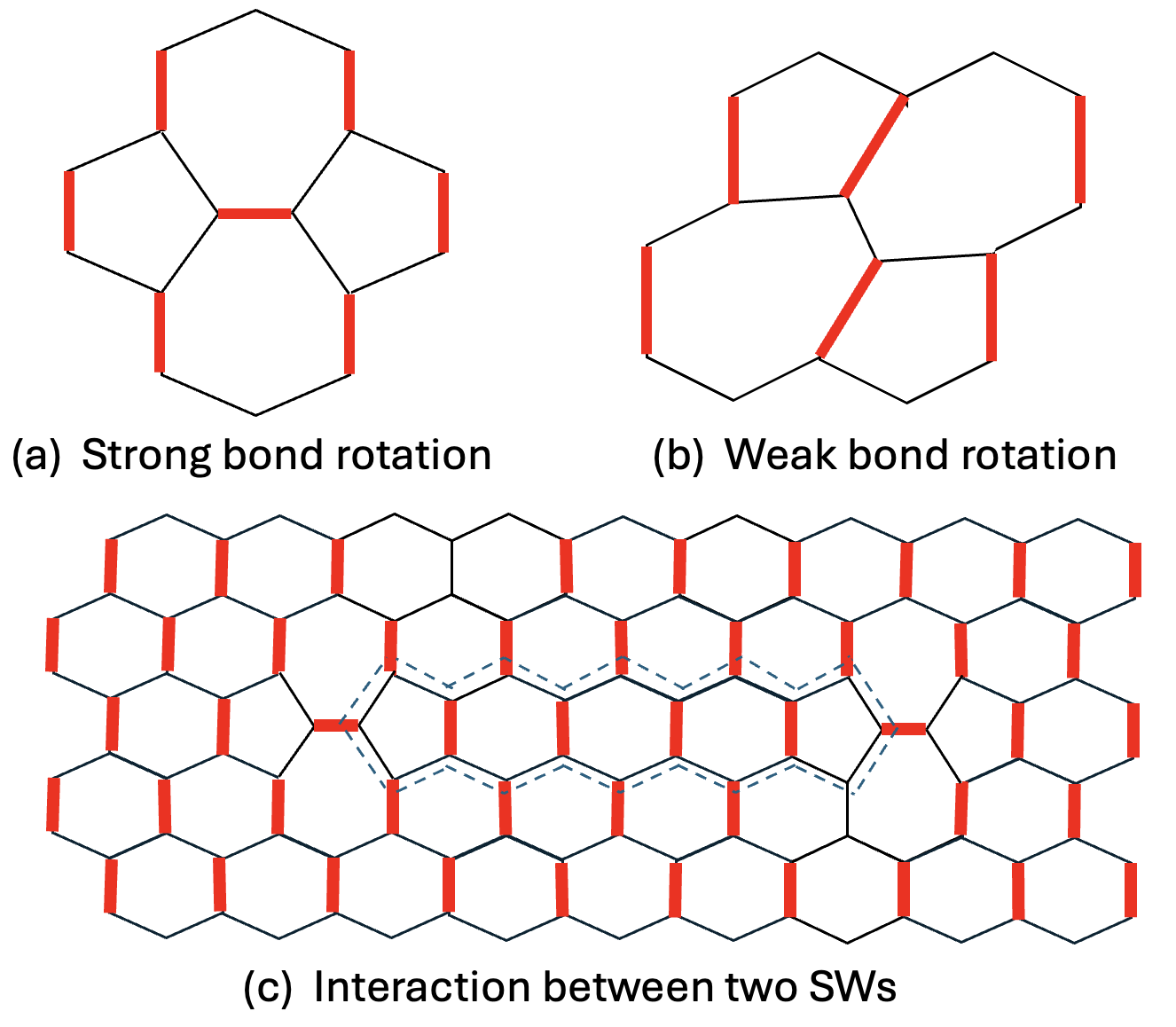}
    \caption{Computation of the flux energies in the gapped Kitaev QSL phase. (a) and (b) show SW defect obtained from a strong and weak bond rotation respectively in the gapped QSL phase. (c)  The loop shown in dashed line is an illustrative example of the loops that we consider to obtain the interaction between two SW defects. }
    \label{fig_perturbation_gapped kitaev}
\end{figure}

To compute the energy of the fluxes and their interaction in the gapped Kitaev phase, we follow Ref.~\cite{petrova_unpaired_2014}. For any loop with $N_b$ total number of bonds, $n_s$ strong bonds and $n_w$ weak bonds and $n_v$ vertices that are connected to strong bonds not lying on the loop, the energy of the loop is given by:
\begin{align}
    E_\bigcirc&=(-1)^{N_b/2}W_\bigcirc\int \frac{d\omega}{2\pi} J^{n_w}
    \left(\frac{2J_{z}}{\omega^2+(2J_{z})^2}\right)^{n_s}\nonumber\\
    &\hspace{4cm}\times\left(\frac{\omega}{\omega^2+(2J_{z})^2}\right)^{n_v}
\end{align}
Here we have chosen a representative point in the gapped phase such that $J_{z}>>J_{x}=J_{y}=J$.  Note that $n_s+n_w=N_b$. Also as described in Ref.~\cite{petrova_unpaired_2014}, only the even sided loops contribute to flux energies, the energies for the odd sided loops cancel because of the $(-1)^{N_b/2}$ factor.

For a single SW, the energy of the leading order energy corrections comes from the loops that encloses a 5-7 and 7-7. 
Also note that in the gapped phase, we can obtain SW defect either by rotating a strong bond or a weak bond. For these two cases, the energy corrections for generic flux sectors are given by,
\begin{align}
    &E_\text{SW}^{(s)}=-\sum_{\text{5-7}}\frac{7J^6}{256 J_{z}^5}W_5W_7+\frac{9J^8}{2048J_z^7}W_7W_{7'}\\
    &E_\text{SW}^{(w)}=-\sum_{\text{5-7}}\frac{5J^8}{2048 J_{z}^7}W_5W_7+\frac{9J^8}{2048J_z^7}W_7W_{7'}
    \label{eq_energy_gapped}
\end{align}
Here superscript $(s)$ and $(w)$ denote SW obtained from strong and weak bond rotations, respectively, and $W_5, W_7$ represent the fluxes on pentagons and heptagons, respectively.
Computing the above expression for Lieb, PT and Uniform-flux, we get for the strong bond rotations:
\begin{align}
    &E_\text{Lieb}^{(s)}=-\frac{7J^6}{64J_z^5}-\frac{9J^8}{2048J_z^7}\\
    &E_\text{PT}^{(s)}=\frac{9J^8}{2048J_z^7}\\
    &E_\text{Uniform}^{(s)}=\frac{7J^6}{64J_z^5}-\frac{9J^8}{2048J_z^7}
    \label{eq_energy_gapped_strong}
\end{align}
and for the weak bond rotation:
\begin{align}
     &E_\text{Lieb}^{(w)}=-\frac{29J^8}{2048J_z^7}\\
    &E_\text{PT}^{(w)}=\frac{9J^8}{2048J_z^7}\\
    &E_\text{Uniform}^{(s)}=\frac{11J^8}{2048J_z^7}
    \label{eq_energy_gapped_weak}
\end{align}
Thus the Lieb flux gets a stabilization energy, hence becomes the ground state of an isolated SW.

For two SWs, there are still two fold degeneracy even if we assign Lieb fluxes on them. Either two SW can be in same (ferromagnetic) $\mu^z$ or opposite (antiferromagnetic) $\mu^z$ state. To show which of them gets becomes the lower energy configuration, it is sufficient to compute the energy of the smallest loop that encloses two pentagons from different SW. Also similar to the single SW case, there might be various combinations to obtain each SW.   However, instead of going into such details, here we try to prove that the interaction is ferromagnetic, and give an estimate of the interaction energy. Note that for distant SWs, the smallest loop that encloses two pentagons has $4n_\text{hexa} +8$ bonds, where $n_\text{hexa}$ is the number of hexagons on the loop. So the interaction energy is always proportional to: $(-1)^{2n_\text{hexa}+4}W_5W_{5'}=W_5W_{5'}$. Clearly, it is minimized by $W_5=W_{5'}$, as they have imaginary fluxes. This establish that the SWs have ferromagnetic interactions. Further to estimate the energy of this interaction note that depending on the orientations and placements of the two SWs, $N_b, n_v, n_s,n_w$ varies giving rise to different integral expressions for Eq.~\ref{eq_energy_gapped}. Here we consider a simple situation when the two SWs are generated by a strong bond rotation and  separated along the parallel direction of the rotated bond as shown by the dashed line in Fig.~\ref{fig_perturbation_gapped kitaev}. For the large separation, the energy of this loop can then be written as,
\begin{align}
    E_{\bigcirc_{55'}}&\approx W_5W_{5'}\int\frac{d\omega}{2\pi} J^{2r}\left(\frac{\omega}{\omega^2+(2J_z)^2}\right)^{2r}\\
    &\sim \frac{J^{2r}}{J_z^{2r-1}}W_5W_{5'}\approx J_{z}\exp\left(-2r \log\left(\frac{J_z}{J}\right)\right)W_5W_{5'}
\end{align}
Note that in the previous expression we have approximated the number of bonds and vertices  on the loops to be $2r$. Here the distance $r$ is a dimensionless number which can be converted into actual separation by a conversion factor involving lattice length scales. If we consider a generic separation the above expression modifies keeping the exponential decay unchanged. The factor in front of $r$ changes depending on the considered loop, and remains anisotropic. So generically the interaction term can be written as
\begin{align}
    E_{\bigcirc_{55'}}\approx e^{-\alpha r}W_5W_{5'}
\end{align}
with $\alpha\propto \log (J_z/J)$.

\end{appendix}

\bibliography{references}

\end{document}